\documentclass[aps,prd,preprintnumbers,showpacs,twocolumn,groupedaddress,nofootinbib]{revtex4}
\usepackage{graphicx}
\usepackage{latexsym}
\usepackage{amsfonts}
\usepackage{amssymb}
\usepackage{amsmath}
\usepackage{slashed}
\usepackage{feynmp}
\def\bea{\begin{eqnarray}}
\def\eea{\end{eqnarray}}
\def\beq{\begin{equation}}
\def\eeq{\end{equation}}
\def\bey{\begin{eqnarray}}
\def\eey{\end{eqnarray}}

\def\lsim{\mathrel{\raise.3ex\hbox{$<$\kern-.75em\lower1ex\hbox{$\sim$}}}}
\def\gsim{\mathrel{\raise.3ex\hbox{$>$\kern-.75em\lower1ex\hbox{$\sim$}}}}

\begin{document}

\title{Phenomenology of Dirac Neutralino Dark Matter}  
\author{Matthew R.~Buckley$^{1}$, Dan Hooper$^{1,2}$, and Jason Kumar$^3$}
\affiliation{$^1$Center for Particle Astrophysics, Fermi National Accelerator Laboratory, Batavia, IL 60510, USA}
\affiliation{$^2$Department of Astronomy and Astrophysics, University of Chicago, Chicago, IL 60637, USA}
\affiliation{$^3$Department of Physics, University of Hawaii, Honolulu, HI  96822, USA}

\date{\today}

\begin{abstract}

In supersymmetric models with an unbroken $R$-symmetry (rather than only $R$-parity), the neutralinos are Dirac fermions rather than Majorana. In this article, we discuss the phenomenology of neutralino dark matter in such models, including the calculation of the thermal relic abundance, and constraints and prospects for direct and indirect searches. Due to the large elastic scattering cross sections with nuclei predicted in $R$-symmetric models, we are forced to consider a neutralino that is predominantly bino, with very little higgsino mixing. We find a large region of parameter space in which bino-like Dirac neutralinos with masses between 10 and 380 GeV can annihilate through slepton exchange to provide a thermal relic abundance in agreement with the observed cosmological density, without relying on coannihilations or resonant annihilations. The signatures for the indirect detection of Dirac neutralinos are very different than predicted in the Majorana case, with annihilations proceeding dominately to $\tau^+ \tau^-$, $\mu^+ \mu^-$ and $e^+ e^-$ final states, without the standard chirality suppression. And unlike Majorana dark matter candidates, Dirac neutralinos experience spin-independent scattering with nuclei through vector couplings (via $Z$ and squark exchange), leading to potentially large rates at direct detection experiments. These and other characteristics make Dirac neutralinos potentially interesting within the context of recent direct and indirect detection anomalies. We also discuss the case in which the introduction of a small Majorana mass term breaks the $R$-symmetry, splitting the Dirac neutralino into a pair of nearly degenerate Majorana states.

\end{abstract}

\pacs{95.35.+d, 14.80.Ly; FERMILAB-PUB-13-256-A, UH511-1213-2013, CETUP2013-002}
\maketitle

\section{Introduction: Dirac Neutralinos as Dark Matter}

In addition to offering a solution to the electroweak hierarchy problem and enabling gauge coupling unification, weak-scale supersymmetry has been motivated by its ability to provide a viable dark matter candidate in the form of the lightest supersymmetric particle (LSP). The lightest neutralino, in particular, has received a great deal of attention within this context~\cite{neutralino}.  In the minimal supersymmetric standard model (MSSM), the lightest neutralino is a Majorana fermion, and is a mixture of the superpartners of the neutral gauge and Higgs bosons. As a consequence of recent null results from the Large Hadron Collider (LHC) and direct dark matter searches, however, much of the parameter space of the MSSM has been ruled out (for recent studies, see Refs.~\cite{Buchmueller:2012hv,blind,Feng:2011aa,Hisano:2012wm,Altmannshofer:2012ks,Feng:2013pwa}). For these and other reasons, an increasing amount of interest has been directed toward alternative realizations of weak-scale supersymmetry. 

Among other possibilities are low-energy supersymmetric models in which the commonly assumed $R$-parity is extended to a symmetry (or, in other words, supersymmetric models in which the underlying $R$-symmetry is not broken to a parity). A number of attractive features can be found in such $R$-symmetric supersymmetric models. Most practically, the LHC's sensitivity to squark production can be significantly reduced in $R$-symmetric models, enabling first and second generation squarks as light as $\sim$700 GeV to have escaped detection~\cite{Kribs:2012gx}.  Furthermore, a variety of flavor observables are much less constraining in $R$-symmetric models than in the MSSM. Whereas in the MSSM, such observations force one to consider supersymmetry breaking scenarios which are approximately minimally flavor violating (a fact known as the supersymmetric flavor problem), order unity flavor violating couplings are allowed in $R$-symmetric models~\cite{Kribs:2007ac,Fok:2010vk}. The degree of electroweak fine tuning required in $R$-symmetric models can also be reduced relative to that found in more traditional supersymmetric frameworks~\cite{Fox:2002bu}.

The phenomenology of neutralino dark matter is very rich and interesting in $R$-symmetric models~\cite{Hsieh:2007wq,Belanger:2009wf,Chun:2009zx}. As a consequence of the $R$-symmetry, gauginos cannot acquire Majorana masses, and thus must instead be Dirac particles. This requires new chiral superfields in adjoint representations of the standard model gauge groups, which combine with the Majorana gauginos to form Dirac states. In terms of annihilation and scattering with nuclei, Dirac particles can behave quite differently than Majorana dark matter candidates.  In particular, whereas the cross section for Majorana fermions annihilating to fermion-antifermion pairs at rest (such as in the halo of the Milky Way) is generically suppressed by a factor of $m^2_f/m^2_{\chi}$, Dirac particles do not experience such chirality suppression~\cite{Harnik:2008uu,Hsieh:2007wq,Belanger:2009wf,Chun:2009zx}. This opens the possibility that the dark matter may be annihilating efficiently to light fermion final states, including $e^+ e^-$, $\mu^+ \mu^-$, or $\nu \bar{\nu}$, with important implications for indirect searches.  Furthermore, unlike Majorana particles, Dirac neutralinos can scatter coherently ({\it i.e.}~through spin-independent interations) with nuclei through vector couplings~\cite{Hsieh:2007wq,Belanger:2009wf,Chun:2009zx}. To evade the constraints from XENON100 and other direct detection experiments, we must suppress the Dirac neutralino's coupling to the $Z$ (by ensuring very little mixing with the higgsinos) {\it and} require that the squarks be quite heavy. This forces us toward a region of parameter space in which the LSP is a highly bino-like Dirac neutralino, annihilating largely through slepton exchange to electrons, muons, taus, and neutrinos. We will show that Dirac binos with masses in the range of approximately 10 GeV to 380 GeV (and higher if neutralino-slepton coannihilations are efficient) can provide a thermal relic abundance that is in agreement with the observed cosmological density of dark matter. 

Much of the dark matter phenomenology described in the previous paragraph can be altered significantly if the $R$-symmetry is even slightly broken, leading to a mass splitting between the two Majorana states which constitute our Dirac neutralino. In this pseudo-Dirac case, the Majorana nature of the LSP is restored for the purposes of indirect detection, once again suppressing the low-velocity annihilation cross section to light fermions.  Furthermore, pseudo-Dirac neutralinos can scatter with nuclei through vector couplings only inelastically, by upscattering the lightest Majorana neutralino into the slightly heavier state. If the mass splitting between these Majorana states is less than $\sim$0.5-5 keV, the dark matter will behave as a Dirac particle for the purposes of direct detection, while for mass splittings larger than $\sim$20-200 keV, the Majorana-like behavior of the MSSM will be restored. In the intermediate range, with mass splittings of $\sim$1-100 keV, the event rates at direct detection experiments will depend sensitively on the mass of the target nuclei and on the velocity of the incoming particle. For roughly GeV-scale mass splittings or less, $\delta m_{\chi} \lsim m_{\chi}/20$, the freeze-out of our dark matter candidate will proceed largely as predicted for a Dirac state, while significantly larger splittings restore the MSSM-like Majorana behavior. 

In this article, we explore the dark matter phenomenology of Dirac and pseudo-Dirac neutralinos. In Sec.~\ref{rsymmetric}, we briefly introduce supersymmetric models with an $R$-symmetry. In Sec.~\ref{directsec} we calculate the elastic scattering cross section of a Dirac neutralino and compare this to the current and projected sensitivities of direct detection experiments. In Sec.~\ref{annihilation}, we calculate the annihilation cross section for a bino-like Dirac neutralino, and evaluate the thermal relic abundance predicted for such a particle. Using the results of that section, we proceed in Sec.~\ref{indirect} to discuss the implications for indirect detection. In Sec.~\ref{anomalies}, we briefly comment on Dirac neutralinos within the context of recent direct and indirect detection anomalies. In Sec.~\ref{pseudodirac}, we extend our discussion to the case of a pseudo-Dirac
neutralino, with small Majorana masses. In Sec.~\ref{summary}, we summarize our results and conclusions.

\section{$R$-Symmetric Supersymmetry}
\label{rsymmetric}

Despite the fact that the supersymmetry algebra explicitly contains a continuous $R$-symmetry, this symmetry is almost universally assumed throughout the literature to be broken down a $Z_2$ parity. This is at least in part because, given the particle content of the MSSM, an unbroken $R$-symmetry forbids masses for both gauginos and higgsinos (each of which carry non-zero $R$ charge), and is thus clearly in conflict with observation. If degrees-of-freedom beyond those described by the MSSM are introduced, however, such obstacles can be circumvented. In particular, although the inclusion of a $\mu$-term is prohibited by the $R$-symmetry, we can still generate masses for the higgsinos if the Higgs sector is enlarged to include the multiplets $R_u$ and $R_d$, each with $R$-charge of $+2$, and which transform in the same way as $H_d$ and $H_u$ under $SU(2)_L \times U(1)_{Y}$~\cite{Kribs:2007ac}. This allows for terms of the form $ \mu_u H_u R_u + \mu_d H_d R_d$ in the $R$-symmetry preserving superpotential. Unlike $H_u$ and $H_d$, $R_u$ and $R_d$ do not participate in electroweak symmetry breaking (they have zero vacuum expectation values), but they do allow for the generation of higgsino masses without breaking the $R$-symmetry (alternatively, see Ref.~\cite{Nelson:2002ca,Davies:2011mp}. We note that within the context of ${\cal N}=2$ supersymmetry, $H_u$ and $R_u$ (or $H_d$ and $R_d$) constitute a complete hypermultiplet.

In order to generate gaugino masses in an $R$-symmetric model, the gauginos must be Dirac fermions. This can be arranged by pairing up the gauginos with additional degrees-of-freedom, such as combining the Majorana gluinos of the MSSM with an additional color octet to yield Dirac gluinos, and combining Majorana binos and winos with a $U(1)_Y$ singlet ($B'$) and a $SU(2)_L$ triplet ($W'$), each with $R=0$, respectively~\cite{Hall:1990hq,Randall:1992cq}. Such additional particle content in the weak-scale spectrum can be motivated in models of ${\cal N}=2$ supersymmetry~\cite{fayet}. In that case, the gauginos and the new adjoint states form a complete ${\cal N}=2$ vector multiplet.

Although a number of seemingly viable models with Dirac gauginos have been proposed in the literature~\cite{Hall:1990hq,Randall:1992cq,Nelson:2002ca,Fox:2002bu,Antoniadis:1992fh,Pomarol:1998sd,Chacko:2004mi,Carpenter:2005tz,Nomura:2005rj,Benakli:2011kz,Benakli:2011vb,Abel:2011dc,Davies:2012vu}, for concreteness we consider a model with the standard kinetic potential and the following superpotential:
\begin{eqnarray}
W &=& \int d^2\theta \bigg[y_u Q U^c H_u + y_d Q D^c H_d + y_e L E^c H_d \nonumber \\
&+& \mu_u H_u R_u + \mu_d H_d R_d + \frac{1}{\sqrt{2}}g' B' (-H_d R_d+H_u R_u) \nonumber \\
&+&\sqrt{2} g W'_i (H_d \sigma_i R_d + H_u \sigma_i R_u) \bigg] + \mbox{h.c.}
\end{eqnarray}
Here, $\sigma_i$ are the $SU(2)_L$ generators, and we have suppressed the flavor structure. The most general $R$-symmetry preserving supersymmetry breaking sector is limited to the following ``supersoft'' terms~\cite{Fox:2002bu}:
%
%In terms of $\mu_u$ and $\mu_d$ appearing in the superpotential, and $m_B$ and $m_T$ appearing in the the following SUSY-%breaking terms:
% 
\begin{eqnarray}
{\cal L}_{\slashed{\rm SUSY}} &=& m_1 \tilde{B}\tilde{B}' + m_2 \tilde{W} \tilde{W}' + m_3 \tilde{G}\tilde{G}' + B_\mu H_u H_d \nonumber \\
&+&\sum_{\rm scalars} m_{\phi}^2 \, \phi^* \phi,
\end{eqnarray}
where the sum runs over all of the scalars in the spectrum, including the new states in the chiral adjoint representations. 

\begin{widetext}
After electroweak symmetry breaking, the Dirac neutralino mass matrix is given by:
%\
%\begin{widetext}
\begin{eqnarray*}
\left(\begin{array}{cccc} \tilde{B}' & \tilde{W}' & \tilde{H}_d & \tilde{H}_u \end{array} \right)\left(\begin{array}{cccc} m_1 & 0 & -m_Z s_W \cos\beta & m_Z s_W \sin\beta \\ 
   0 & m_2 & m_Z c_W \cos \beta & -m_Z c_W \sin\beta \\
   -m_Z s_W \cos\beta & m_Z c_W \cos\beta & -\mu_d & 0 \\
   m_Z s_W \sin\beta & -m_Z c_W \sin \beta & 0 & -\mu_u \end{array}\right)\left(\begin{array}{c} \tilde{B} \\ \tilde{W} \\ \tilde{R}_d \\ \tilde{R}_u \end{array}\right) ,
\end{eqnarray*} 
\end{widetext}
where $s_W$ and $c_W$ are the sine and cosine of the Weinberg angle. Although this matrix is similar to that found for Majorana neutralinos in the MSSM, notice that the $\mu_d$ and $\mu_u$ terms now each appear on diagonal entries.  

In additional to Dirac gaugino masses and the extended Higgs sector, we note that both $A$-terms and $\mu$-terms are forbidden in $R$-symmetric supersymmetry models and thus there is no left-right sfermion mixing. In the following four sections, we will consider the case in which the $R$-symmetry is unbroken, and then extend our discussion in Sec.~\ref{pseudodirac} to include the possibility of a broken $R$-symmetry with non-zero Majorana mass terms.

\section{Elastic Scattering and Direct Detection}
\label{directsec}

In the MSSM, Majorana neutralinos undergo spin-independent (coherent) scattering with nuclei via both scalar Higgs and squark exchange, and spin-dependent scattering through exchange of the $Z$. In the case of a Dirac neutralino, however, the vector interaction of the $Z$ exchange instead leads to a spin-independent interaction. As we will see, the elastic scattering cross section induced by this process (as well as by squark exchange) can be quite large, leading to significant constraints from direct detection experiments.

\subsection{$Z$ Exchange}

\begin{figure*}
\includegraphics[width=4.8in]{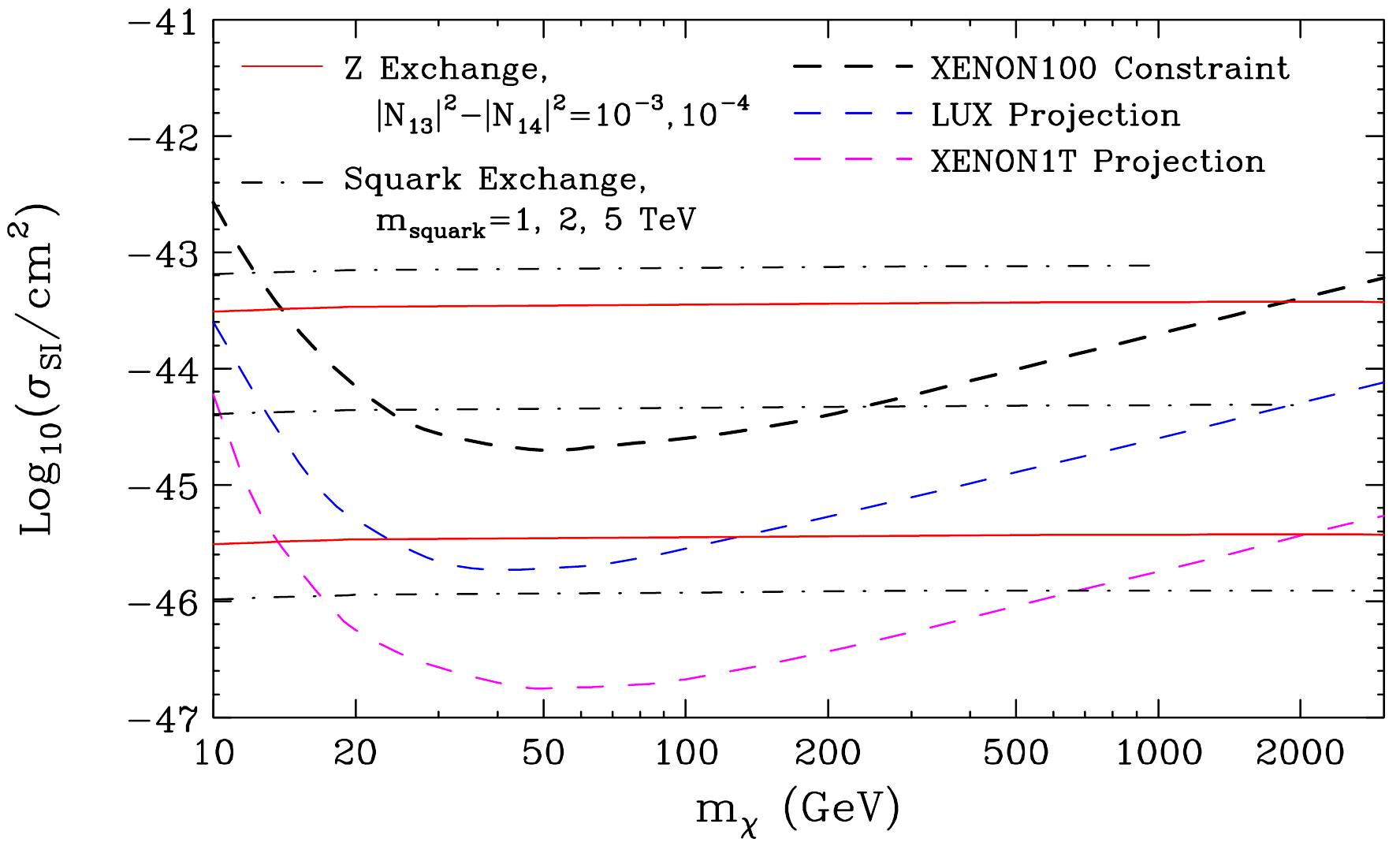}
\caption{The elastic scattering cross section per nucleon for a Dirac neutralino with a xenon target, due to $Z$-exchange and squark exchange processes. Results are shown for two choices of the higgsino content ($|N_{13}|^2-|N_{14}|^2=10^{-3}$ and $10^{-4}$, upper and lower solid red lines, respectively) and three choices of the squark mass ($m_{\tilde q}=1$, 2, and 5 TeV, from top-to-bottom). For $m_{\chi} \gsim 20$ GeV, present constraints from XENON100~\cite{xenon} exclude a Dirac neutralino dark matter candidate unless the magnitude of its higgsino component is very small (corresponding to $\mu_{u,d} \times (100 \, {\rm GeV}/m_{\chi}) \gsim 2$ TeV) {\it and} the squarks are quite heavy ($\gsim$2 TeV). The projected sensitivity of LUX~\cite{lux} and XENON1T~\cite{xenon1t} will extend this reach to values of $\mu_{u,d}$ and $m_{\tilde q}$ as high as $\sim$5 TeV. See text for details.} 
\label{direct}
\end{figure*}

The spin-independent scattering cross section can be written in terms of the spin-averaged squared
matrix element:
\bea
\sigma_{SI} = {\mu^2 \over 16\pi m_{\chi}^2 m_A^2 } \left({1\over 4} \sum_{\rm spins} |{\cal M}|^2 \right),
\eea
where $\mu$ is the reduced mass and $m_A$ is the mass of the target nucleus. The effective operator for dark matter scattering through $Z$-exchange can be written as:
\bea
{\cal O}_Z &=& \lambda_{\chi V} \lambda_{qV}(1/m_Z^2) (\bar \chi \gamma^\mu \chi)(\bar q \gamma_\mu q) \nonumber \\
&+&\lambda_{\chi V} \lambda_{qA}(1/m_Z^2) (\bar \chi \gamma^\mu \chi)(\bar q \gamma_\mu \gamma^5 q)
\nonumber\\
&+&\lambda_{\chi A} \lambda_{qV}(1/m_Z^2) (\bar \chi \gamma^\mu \gamma^5 \chi)(\bar q \gamma_\mu q) \nonumber \\
&+&\lambda_{\chi A} \lambda_{qA}(1/m_Z^2) (\bar \chi \gamma^\mu \gamma^5 \chi)(\bar q \gamma_\mu \gamma^5 q),
\label{zoperator}
\eea
where 
\bea
\lambda_{qV} &=& {g \over \cos \theta_W}[T^3_{qL} - 2Q_{q} \sin^2 \theta_W ]
\nonumber\\
\lambda_{qA} &=& {g \over \cos \theta_W}[-T^3_{qL} ]
\nonumber\\
\lambda_{\chi V} &=& {g \over  \cos \theta_W}[T^3_{\chi R} +T^3_{\chi L} ]
\nonumber\\
\lambda_{\chi A} &=& {g \over \cos \theta_W}[T^3_{\chi R} -T^3_{\chi L} ].
\eea
Only the first term in Eq.~\eqref{zoperator} contributes to spin-independent scattering in the low-velocity limit. This yields a squared matrix element (spin-averaged) given by:
\bea
&&{1\over 4}\sum_{\rm spins} |{\cal M}_{SI}|^2 \\
&\simeq& 16 {m_{\chi}^2 m_A^2 \over m_Z^4} \lambda_{\chi V}^2 \left[ \sum_q \lambda_{qV} 
[Z  B_{qV}^p + (A-Z) B_{qV}^n] \right]^2, \nonumber
\eea
where $B_{uV}^p = B_{dV}^n = 2$ and $B_{uV}^n = B_{dV}^p = 1$. $Z$ and $A$ correspond to the atomic number and atomic mass of the target nucleus. This leads to a $Z$-induced spin-independent cross section with protons and neutrons given by:
\bea
\sigma^{(Z)}_{\chi p} &\simeq& \frac{\mu^2 g^4 \, (0.5-2 \sin^2 \theta_{W})^2}{\pi \cos^4 \theta_W m^4_Z} \bigg[|N_{13}|^2 - |N_{14}|^2 \bigg]^2 \\
& \simeq & 6.0 \times 10^{-44} \, {\rm cm}^2 \, \times \bigg(\frac{\mu}{m_p}\bigg)^2 \bigg( \frac{|N_{13}|^2 - |N_{14}|^2}{0.01} \bigg)^2 \nonumber
\eea
and
\bea
\sigma^{(Z)}_{\chi n} &\simeq& \frac{\mu^2 g^4}{4 \pi \cos^4 \theta_W m^4_Z} \bigg[|N_{13}|^2 - |N_{14}|^2 \bigg]^2 \\
& \simeq & 1.0 \times 10^{-41} \, {\rm cm}^2 \, \times \bigg(\frac{\mu}{m_n}\bigg)^2 \bigg( \frac{|N_{13}|^2 - |N_{14}|^2}{0.01} \bigg)^2, \nonumber
\eea
where $N_{ij}$ are elements of the matrix the diagonalizes the neutralino mass matrix given in Sec.~\ref{rsymmetric}. The quantities $|N_{13}|^2$ and $|N_{14}|^2$ describe the fraction of the lightest neutralino's composition that is made up of higgsinos.

These cross sections are quite large, and are highly constrained by existing direct detection experiments. In Fig.~\ref{direct}, we plot this contribution to the spin-independent elastic scattering cross section (per nucleon, for the case of a xenon target) as red solid lines, for values corresponding to $|N_{13}|^2-|N_{14}|^2=10^{-3}$ (upper) and $10^{-4}$ (lower). The current constraint from the XENON100 experiment~\cite{xenon} already excludes dark matter in the form of a Dirac neutralino unless the magnitude of its higgsino component is very small (corresponding roughly to $\mu_{u,d} \times (100 \, {\rm GeV}/m_{\chi}) \gsim 2$ TeV), or the neutralino is very light. In the future, experiments such as LUX~\cite{lux} and XENON1T~\cite{xenon1t} will become even more sensitive to Dirac neutralinos and other Dirac dark matter candidates with non-zero couplings to the $Z$. Neutrino detectors looking for annihilation of dark matter in the Sun also place limits on the annihilation channels \cite{Boliev:2013ai}, however, for the models presented in this paper, the direct detection limits are more stringent.

\subsection{Squark Exchange}

In a case in which the Dirac neutralino has very little higgsino content, thus suppressing its coupling to the $Z$, one still must be mindful of the potentially large spin-independent scattering cross section induced by squark exchange. Unlike in the case of a Majorana neutralino, the process of a Dirac particle scattering via squark exchange does not require a spin-flip, thus avoiding the $m^2_q$ suppression exhibited in the MSSM with minimal flavor violation. 

The Lagrangian for the bino-quark-squark interaction is given by
\begin{equation}
{\cal L} \supseteq \sqrt{2} g' \, Y_q (\chi P_{L/R} \bar{q}) {\tilde{q}}_{R/L} +\mbox{h.c.},
\end{equation}
where $Y_q$ is the hypercharge of the quark and $P_{R/L}$ is the right- or left-projection operator. 
%So, over the first generation squarks, we have:
%\begin{widetext}
%\begin{equation}
%{\cal L} \supseteq \sqrt{2} g'\left(+\frac{1}{6} \right) \bar{\chi} P_L u  \bar{\tilde{u}}_L+\sqrt{2} g' \left(+\frac{1}{6} \right) \bar{\chi} P_L d  %\bar{\tilde{d}}_L +\sqrt{2} g'\left(+\frac{2}{3} \right) \bar{\chi} P_L u  \bar{\tilde{u}}_L+\sqrt{2} g' \left(-\frac{1}{3} \right) \bar{\chi} P_L d  %\bar{\tilde{d}}_L +\mbox{h.c.}
%\end{equation}
%\end{widetext}
%
%Therefore, 
%Assuming four different squark mass eigenstates $m_{\tilde{u}_L}$, $m_{\tilde{d}_L}$, $m_{\tilde{u}_R}$, and $m_{\tilde{d}_R}$, t
The effective operator for a light quark/bino scattering interaction is then
\begin{equation}
{\cal L}_{\rm eff} = \frac{2 g'^2 Y_q^2}{m_{\tilde{q}}^2} \left[ \bar{\chi} P_{R/L} q \right] \left[  \bar{q} P_{L/R} \chi \right].
\end{equation}
Expanding out the 4-fermion operator (assuming $P_R/P_L$ for the moment), and applying the Fierz transformations~\cite{Nieves:2003in}, we arrive at:
\begin{widetext}
\begin{eqnarray}
\left[ \bar{\chi} P_{R} q \right] \left[  \bar{q} P_{L} \chi \right] & = & \frac{1}{4} \left( [ \bar{\chi} q][  \bar{q} \chi] +[ \bar{\chi} \gamma^5 q][  \bar{q} \chi] -[ \bar{\chi} q][  \bar{q} \gamma^5 \chi]  -[ \bar{\chi}\gamma^5 q][  \bar{q}\gamma^5 \chi] \right) \\
% & = & \frac{1}{16} \times \left( [\bar{q} q][  \bar{\chi} \chi] +[\bar{q} \gamma^\mu q][  \bar{\chi} \gamma_\mu\chi] +\frac{1}{2} [\bar{q} \sigma^{\mu\nu}q][  \bar{\chi} \sigma_{\mu\nu} \chi] -[\bar{q} \gamma^\mu\gamma^5  q][  \bar{\chi}\gamma_\mu\gamma^5 \chi] +[\bar{q}\gamma^5 q][  \bar{\chi} \gamma^5\chi]  \right. \nonumber \\
% & &  +[\bar{q} \gamma^5 q][  \bar{\chi} \chi] +[\bar{q} \gamma^\mu\gamma^5 q][  \bar{\chi} \gamma_\mu\chi] +\frac{1}{2} [\bar{q} \sigma^{\mu\nu}q][  \bar{\chi} \sigma_{\mu\nu} \chi] -[\bar{q} \gamma^\mu q][  \bar{\chi}\gamma_\mu\gamma^5 \chi] +[\bar{q} q][  \bar{\chi} \gamma^5\chi]  \nonumber \\
% & &  -[\bar{q}  q][  \bar{\chi} \gamma^5 \chi] -[\bar{q} \gamma^\mu q][  \bar{\chi} \gamma_\mu\gamma^5 \chi] -\frac{1}{2} [\bar{q} \sigma^{\mu\nu}q][  \bar{\chi} \sigma_{\mu\nu} \chi] +[\bar{q} \gamma^\mu\gamma^5 q][  \bar{\chi}\gamma_\mu \chi] -[\bar{q}\gamma^5  q][  \bar{\chi}\chi]  \nonumber \\
% & & \left. -[\bar{q} q][  \bar{\chi} \chi] +[\bar{q} \gamma^\mu q][  \bar{\chi} \gamma_\mu\chi] -\frac{1}{2} [\bar{q} \sigma^{\mu\nu}q][  \bar{\chi} \sigma_{\mu\nu} \chi] -[\bar{q} \gamma^\mu\gamma^5  q][  \bar{\chi}\gamma_\mu\gamma^5 \chi] -[\bar{q}\gamma^5 q][  \bar{\chi} \gamma^5\chi]  \right) \nonumber \\
% & = & \frac{1}{8}\times \left( [\bar{q} \gamma^\mu q][  \bar{\chi} \gamma_\mu \chi] +[\bar{q} \gamma^\mu\gamma^5  q][  \bar{\chi} \gamma_\mu\chi] - [\bar{q} \gamma^\mu  q][  \bar{\chi} \gamma_\mu \gamma^5\chi] - [\bar{q} \gamma^\mu\gamma^5  q][  \bar{\chi} \gamma_\mu \gamma^5\chi]  \right) \nonumber \\
 & = & \frac{1}{2} [\bar{q} \gamma^\mu P_R q] [\bar{\chi} \gamma_\mu P_L \chi]. \nonumber
\end{eqnarray}
\end{widetext}
Thus, the Lagrangian for this interaction is
\begin{equation}
{\cal L}_{\rm eff} = \frac{g'^2 Y_q^2}{m_{\tilde{q}}^2} \left [\bar{q} \gamma^\mu P_{L/R} q \right] \left[\bar{\chi} \gamma_\mu P_{R/L} \chi\right],
\end{equation}
and the spin-independent interaction is determined by the vector piece:
\begin{equation}
{\cal L}_{\rm SI} = 
\left( {g'^2 Y_{q_L}^2 \over 4m_{\tilde q_L}^2 } + {g'^2 Y_{q_R}^2 \over 4m_{\tilde q_R}^2 } \right) \left [\bar{q} \gamma^\mu  q \right] \left[\bar{\chi} \gamma_\mu  \chi\right].
\end{equation}
For a vector interaction, the squark-induced cross section for protons/neutrons is given by:
\begin{equation}
\sigma_{p,n}  =  \frac{\mu^2}{\pi} f_{p,n}^2, 
\end{equation}
where
\begin{eqnarray}
f_p & = & 2\times \left(\frac{Y_{u_L}^2}{4m_{\tilde{u}_L}^2} +\frac{Y_{u_R}^2}{4m_{\tilde{u}_R}^2} \right)+ \left(\frac{Y_{d_L}^2}{4m_{\tilde{d}_L}^2} +\frac{Y_{d_R}^2}{4m_{\tilde{d}_R}^2} \right) \\
f_n & = & \left(\frac{Y_{u_L}^2}{4m_{\tilde{u}_L}^2} +\frac{Y_{u_R}^2}{4m_{\tilde{u}_R}^2} \right)+2\times \ \left(\frac{Y_{d_L}^2}{4m_{\tilde{d}_L}^2} +\frac{Y_{d_R}^2}{4m_{\tilde{d}_R}^2} \right). \nonumber
\end{eqnarray}
Assuming a common squark mass, $m_{\tilde q}$, this yields a spin-independent elastic scattering cross section given by:
\begin{eqnarray*}
\sigma^{({\tilde q})}_{\chi p} & = & \frac{g'^4\mu^2}{4\pi m_{\tilde q}^4} \left[\left(\frac{1}{36}+\frac{4}{9} \right) +\frac{1}{2}\left(\frac{1}{36}+\frac{1}{9} \right) \right]^2\\
 & \approx & \left(1.1 \times 10^{-43}~\mbox{cm}^{2}\right)\left(\frac{\mu}{m_p}\right)^2\left(\frac{1~\mbox{TeV}}{m_{\tilde q}}\right)^4
\end{eqnarray*}
and 
\begin{eqnarray*}
\sigma^{({\tilde q})}_{\chi n} & = & \frac{g'^4\mu^2}{4\pi m_{\tilde q}^4} \left[\frac{1}{2}\left(\frac{1}{36}+\frac{4}{9} \right) +\left(\frac{1}{36}+\frac{1}{9} \right) \right]^2\\
 & \approx & \left(5.5 \times 10^{-44}~\mbox{cm}^{2}\right)\left(\frac{\mu}{m_n}\right)^2\left(\frac{1~\mbox{TeV}}{m_{\tilde q}}\right)^4.
\end{eqnarray*}

This result is shown for the case of a xenon target in Fig.~\ref{direct}. Present constraints require $m_{\tilde q} \gsim 1.5-2.2$ TeV, depending on the mass of the bino.  Taken together with the contribution from $Z$-exchange, we find that in order for a Dirac neutralino to evade existing constraints from XENON100 and other direct detection experiments, the neutralino must possess very little higgsino content (be almost a pure gaugino) {\it and} the squark masses must be quite heavy (unless the neutralino is rather light, $m_{\chi} \lsim 20$ GeV).

Since the $Z$- and squark-exchange processes each contribute to an effective operator of the form $\bar \chi \gamma ^\mu \chi
\bar q \gamma_\mu q$, they will interfere.  As the interference terms will be important only if the amplitudes for these processes are coincidentally similar, we do not calculate them explicitly. Furthermore, it is  not possible for significant destructive interference to occur for interactions with both protons and neutrons.  

\subsection{Higgs Exchange}

For Higgs exchange, the effective scattering operator can be written as
\bea
{\cal O}_{h} &=& \lambda_{\chi h} \lambda_{qh}(1/m_h^2) (\bar \chi \chi)(\bar q  q)
\eea
where $\lambda_{qh} = m_q/ v$ and $v \simeq 246$ GeV is the vacuum expectation value of the Higgs field.

The squared matrix element (spin-averaged) is then given by
\bea
&&{1\over 4}\sum_{spins} |{\cal M}_{SI}|^2 \\
&\simeq& 16 {m_{\chi}^2 m_A^2\over m_h^4} \lambda_{\chi h}^2 \left[ \sum_q \lambda_{qh} [Z  B_{qS}^p + (A-Z) B_{qS}^n] \right]^2, \nonumber
\eea
where $B_{uS}^p = B_{dS}^n \sim 6$ and $B_{uS}^n = B_{dS}^n \sim 4$. This contribution will almost invariably be subdominant to those from $Z$ and squark exchange.

In the next section, we will consider the annihilation cross section for a Dirac neutralino and calculate the thermal relic abundance of such a dark matter candidate. We focus in particular on those scenarios found in this section to be consistent with existing direct detection constraints.

\section{Annihilation and Relic Abundance}
\label{annihilation}

Under standard cosmological assumptions, a single particle species with a $\sim$GeV-TeV scale mass will freeze-out of thermal equilibrium with a relic abundance approximately given by:
\begin{eqnarray}
\Omega_{\chi} h^2 &\approx& \frac{1.04\times10^9}{M_{\rm Pl}} \frac{x_F}{\sqrt{g_{\star}}} \frac{1}{\langle\sigma v\rangle} \nonumber \\
&\approx& 0.12 \, \bigg(\frac{3\times 10^{-26} \, {\rm cm}^3/{\rm s}}{\langle\sigma v\rangle}\bigg), 
\end{eqnarray}
where $M_{\rm Pl}$ is the Planck mass, $x_F =  m_{\chi}/T_{\rm FO}$ is the ratio of the neutralino mass to the freeze-out temperature, $g_{\star}$ is the number of degrees of freedom at the temperature of freeze-out, and $\langle\sigma v\rangle$ is the thermally averaged annihilation cross section evaluated at the temperature of freeze-out. Writing the annihilation cross section as an expansion in powers of velocity, $\sigma v = a + b v^2 + \mathcal{O}(v^4)$, the thermal average at freeze-out is well approximated by $\langle\sigma v\rangle \simeq a+3b/x_F$ ($x_F \approx 20$ for typical weakly interacting massive particles).  For an up-to-date and detailed treatment of dark matter freeze-out, see Ref.~\cite{Steigman:2012nb}. 

For a Dirac neutralino, we can write the effective annihilation cross section in terms of the annihilation cross sections between the two degenerate Majorana states~\cite{Servant:2002aq}:
\begin{equation}
\sigma_{\rm Eff} = \frac{1}{4}\,\sigma_{11} + \frac{1}{2}\,\sigma_{12} + \frac{1}{4}\,\sigma_{22}  
\label{effdirac}
\end{equation}
where $\sigma_{ij}$ denotes the annihilation (or coannihilation) cross section between Majorana mass eigenstates $i$ and $j$.

Given the constraints found in the last section, we focus here on the case of a Dirac neutralino with very little higgsino content; in particular a nearly-pure bino (although a Dirac wino LSP is also a possibility, its annihilation cross section is too large to avoid being underproduced in the early Universe, unless very heavy). For a Dirac bino, the process of neutralino annihilation is dominated by $t$-channel sfermion exchange. Once the direct detection constraint on the squark masses is taken into account, we find that the annihilations must proceed largely to lepton pairs, via slepton exchange.

To determine the thermal relic abundance of dark matter in this scenario, we calculate the relevant cross sections. Like-type annihilations (1-1 or 2-2) are very similar to the standard MSSM-like case, with matrix elements given by:
\begin{widetext}
\begin{eqnarray}
{\cal M}_{11} & = & -\frac{1}{4} \lambda_L^2 A_1\left(\frac{\cos^2\alpha}{M_1^2 - t}+\frac{\sin^2\alpha}{M_2^2 - t}-\frac{\cos^2\alpha}{M_1^2 -u}-\frac{\sin^2\alpha}{M_2^2 - u} \right) [\bar{v}(p_2)\gamma^\mu u(p_1)][ \bar{u}(k_1) \gamma_\mu P_L v(k_2)] \\
& & -\frac{1}{4} \lambda_L^2 A_1 \left(\frac{\cos^2\alpha}{M_1^2 - t}+\frac{\sin^2\alpha}{M_2^2 - t}+\frac{\cos^2\alpha}{M_1^2 -u}+\frac{\sin^2\alpha}{M_2^2 - u} \right) [\bar{v}(p_2)\gamma^\mu\gamma^5 u(p_1)][ \bar{u}(k_1) \gamma_\mu P_L v(k_2)] \nonumber \\
& & -\frac{1}{4} \lambda_R^2 A_2 \left(\frac{\sin^2\alpha}{M_1^2 - t}+\frac{\cos^2\alpha}{M_2^2 - t}-\frac{\sin^2\alpha}{M_1^2 -u}-\frac{\cos^2\alpha}{M_2^2 - u} \right) [\bar{v}(p_2)\gamma^\mu u(p_1)][ \bar{u}(k_1) \gamma_\mu P_R v(k_2)] \nonumber \\
& & +\frac{1}{4} \lambda_R^2 A_2 \left(\frac{\sin^2\alpha}{M_1^2 - t}+\frac{\cos^2\alpha}{M_2^2 - t}+\frac{\sin^2\alpha}{M_1^2 -u}+\frac{\cos^2\alpha}{M_2^2 - u} \right) [\bar{v}(p_2)\gamma^\mu\gamma^5 u(p_1)][ \bar{u}(k_1) \gamma_\mu P_R v(k_2)] \nonumber \\
& & +\frac{1}{2} \lambda_L\lambda_R A_3 \sin\alpha\cos\alpha\left(\frac{1}{M_1^2-t}-\frac{1}{M_2^2-t}+\frac{1}{M_1^2-u}-\frac{1}{M_2^2-u} \right) [\bar{v}(p_2) P_R  u(p_1)][ \bar{u}(k_1) P_R v(k_2)] \nonumber \\
& & +\frac{1}{2} \lambda_L\lambda_R A_3 \sin\alpha\cos\alpha\left(\frac{1}{M_1^2-t}-\frac{1}{M_2^2-t}+\frac{1}{M_1^2-u}-\frac{1}{M_2^2-u} \right) [\bar{v}(p_2) P_L  u(p_1)][ \bar{u}(k_1) P_L v(k_2)] \nonumber \\
& & +\frac{1}{8} \lambda_L\lambda_R A_3 \sin\alpha\cos\alpha\left(\frac{1}{M_1^2-t}-\frac{1}{M_2^2-t}-\frac{1}{M_1^2-u}+\frac{1}{M_2^2-u} \right) [\bar{v}(p_2) \sigma^{\mu\nu} u(p_1)][ \bar{u}(k_1) \sigma_{\mu\nu}P_R v(k_2)] \nonumber \\
& & +\frac{1}{8} \lambda_L\lambda_R A_3 \sin\alpha\cos\alpha\left(\frac{1}{M_1^2-t}-\frac{1}{M_2^2-t}-\frac{1}{M_1^2-u}+\frac{1}{M_2^2-u} \right) [\bar{v}(p_2) \sigma^{\mu\nu} u(p_1)][ \bar{u}(k_1) \sigma_{\mu\nu}P_L v(k_2)],\nonumber
\end{eqnarray}
where
\begin{eqnarray*}
\lambda_L & = & -\frac{g'}{\sqrt{2}}  \\
\lambda_R & = & -\sqrt{2} g'  \\
A_1 & = & (\cos \theta^* - A_L \sin \theta^*)^2 \\
A_2 & = & (\cos \theta^* - A_R \sin \theta^*)^2 \\
A_3 & = & (\cos \theta^* - A_L \sin \theta^*)(\cos \theta^* - A_R \sin \theta^*) \\
t & = & (p_1 - k_1)^2 = m_{\chi^1}^2 + m_f^2 -\frac{s}{2}\left(1-\sqrt{1-\frac{4m_f^2}{s}}\sqrt{1-\frac{4m_{\chi^1}^2}{s}}\cos\theta \right) \\
u & = & (p_1 - k_2)^2 = m_{\chi^1}^2 + m_f^2 +\frac{s}{2}\left(1-\sqrt{1-\frac{4m_f^2}{s}}\sqrt{1-\frac{4m_{\chi^1}^2}{s}}\cos\theta \right).
\end{eqnarray*}
\end{widetext}
In the above, $M_1$ and $M_2$ are the mass eigenstates of the exchanged sleptons or squarks (not to be confused with the bino or wino masses, $m_1$ and $m_2$), and $\alpha$ is the mixing angle between those two states (when $\cos \alpha=1$, $M_1$ and $M_2$ are the masses of the left-handed and right-handed sfermions, respectively). $m_{\chi^1}$ is the mass of the lightest neutralino. In Sec.~\ref{pseudodirac}, we will consider the case in which the $R$-symmetry is broken, splitting the Dirac neutralino into two quasi-degenerate Majorana states, with masses $m_{\chi^1}$ and $m_{\chi^2}$. $\theta$ is the physical angle between the incoming dark matter and the outgoing fermions, not to be confused with the mixing angle $\theta^*$ which mixes the standard Majorana bino with the ``right-handed bino'', which has couplings of $-\sqrt{2} g' \tfrac{1}{2} A_L$ and $-\sqrt{2} g' A_R$ to left- and right-handed sfermions. In the case of ${\cal N}$=2 supersymmetry, $A_R$ and $A_L $ are each set to zero. The standard (MSSM-like) case is recovered for $\sin \theta^*=1$, while the Dirac case corresponds to $\sin \theta^*=1/\sqrt{2}$. The matrix element for 2-2 scattering, ${\cal M}_{22}$, is the same as for ${\cal M}_{11}$, after making the substitutions $\cos \theta^* \rightarrow \sin \theta^*$, $\sin \theta^* \rightarrow -\cos \theta^*$, and $m_{\chi^1} \leftrightarrow m_{\chi^2}$.

%\begin{eqnarray*}
%lambda_L & = & \sqrt{2} g'\left(-\frac{1}{2}\right) (\sin \theta^* + \cos \theta^* A_L) \\
%\lambda_R & = & \sqrt{2} g'\left(+1\right) (\sin \theta^*+ \cos\theta^* A_R)
%\end{eqnarray*}

The matrix element for the coannihilation between the two Majorana states is given by:
\begin{widetext}
\begin{eqnarray*}
{\cal M}_{12} & = & -\frac{1}{2} \lambda_L^2 B_1 \left(\frac{\cos^2\alpha}{M_1^2 - t}+\frac{\sin^2\alpha}{M_2^2 - t}+\frac{\cos^2\alpha}{M_1^2 -u}+\frac{\sin^2\alpha}{M_2^2 - u} \right) [\bar{v}(p_2)\gamma^\mu u(p_1)][ \bar{u}(k_1) \gamma_\mu P_L v(k_2)] \\
& & -\frac{1}{2} \lambda_L^2 B_1 \left(\frac{\cos^2\alpha}{M_1^2 - t}+\frac{\sin^2\alpha}{M_2^2 - t}-\frac{\cos^2\alpha}{M_1^2 -u}-\frac{\sin^2\alpha}{M_2^2 - u} \right) [\bar{v}(p_2)\gamma^\mu\gamma^5 u(p_1)][ \bar{u}(k_1) \gamma_\mu P_L v(k_2)] \\
& & -\frac{1}{2} \lambda_R^2 B_2 \left(\frac{\sin^2\alpha}{M_1^2 - t}+\frac{\cos^2\alpha}{M_2^2 - t}+\frac{\sin^2\alpha}{M_1^2 -u}+\frac{\cos^2\alpha}{M_2^2 - u} \right) [\bar{v}(p_2)\gamma^\mu u(p_1)][ \bar{u}(k_1) \gamma_\mu P_R v(k_2)] \\
& & +\frac{1}{2} \lambda_R^2 B_2 \left(\frac{\sin^2\alpha}{M_1^2 - t}+\frac{\cos^2\alpha}{M_2^2 - t}-\frac{\sin^2\alpha}{M_1^2 -u}-\frac{\cos^2\alpha}{M_2^2 - u} \right) [\bar{v}(p_2)\gamma^\mu\gamma^5 u(p_1)][ \bar{u}(k_1) \gamma_\mu P_R v(k_2)] \\
& & + \lambda_L\lambda_R \sin\alpha\cos\alpha\left(\frac{B_3}{M_1^2-t}-\frac{B_3}{M_2^2-t}-\frac{B_4}{M_1^2-u}+\frac{B_4}{M_2^2-u} \right) [\bar{v}(p_2) P_R  u(p_1)][ \bar{u}(k_1) P_R v(k_2)] \\
& & + \lambda_L\lambda_R \sin\alpha\cos\alpha\left(\frac{B_4}{M_1^2-t}-\frac{B_4}{M_2^2-t}-\frac{B_3}{M_1^2-u}+\frac{B_3}{M_2^2-u} \right) [\bar{v}(p_2) P_L  u(p_1)][ \bar{u}(k_1) P_L v(k_2)] \\
& & +\frac{1}{4} \lambda_L\lambda_R \sin\alpha\cos\alpha\left(\frac{B_3}{M_1^2-t}-\frac{B_3}{M_2^2-t}+\frac{B_4}{M_1^2-u}-\frac{B_4}{M_2^2-u} \right) [\bar{v}(p_2) \sigma^{\mu\nu} u(p_1)][ \bar{u}(k_1) \sigma_{\mu\nu}P_R v(k_2)] \\
& & +\frac{1}{4} \lambda_L\lambda_R \sin\alpha\cos\alpha\left(\frac{B_4}{M_1^2-t}-\frac{B_4}{M_2^2-t}+\frac{B_3}{M_1^2-u}-\frac{B_3}{M_2^2-u} \right) [\bar{v}(p_2) \sigma^{\mu\nu} u(p_1)][ \bar{u}(k_1) \sigma_{\mu\nu}P_L v(k_2)]
\end{eqnarray*}
where
\begin{eqnarray*}
B_1 & = & (\cos\theta^*-A_L\sin\theta^*)(\sin\theta^*+A_L\cos\theta^*) \\
B_2 & = &  (\cos\theta^*-A_R\sin\theta^*)(\sin\theta^*+A_R\cos\theta^*) \\
B_3 & = &  (\cos\theta^*-A_L\sin\theta^*)(\sin\theta^*+A_R\cos\theta^*)\\
B_4 & = &  (\cos\theta^*-A_R\sin\theta^*)(\sin\theta^*+A_L\cos\theta^*)\\
t & = & (p_1 - k_1)^2 = m_f^2 + m_{\chi^1}^2 -\frac{1}{2}\left(s+m_{\chi^1}^2+m_{\chi^2}^2-\sqrt{1-\frac{4m_f^2}{s}}\sqrt{s(s-2(m_{\chi^1}^2+m_{\chi^2}^2)+(m_{\chi^1}^2-m_{\chi^2}^2)^2} \cos\theta \right)\\
u & = & (p_1 - k_2)^2 = m_f^2 + m_{\chi^1}^2 -\frac{1}{2}\left(s+m_{\chi^1}^2+m_{\chi^2}^2+\sqrt{1-\frac{4m_f^2}{s}}\sqrt{s(s-2(m_{\chi^1}^2+m_{\chi^2}^2)+(m_{\chi^1}^2-m_{\chi^2}^2)^2} \cos\theta \right).
\end{eqnarray*}
\end{widetext}

\begin{figure*}
\includegraphics[width=2.9in]{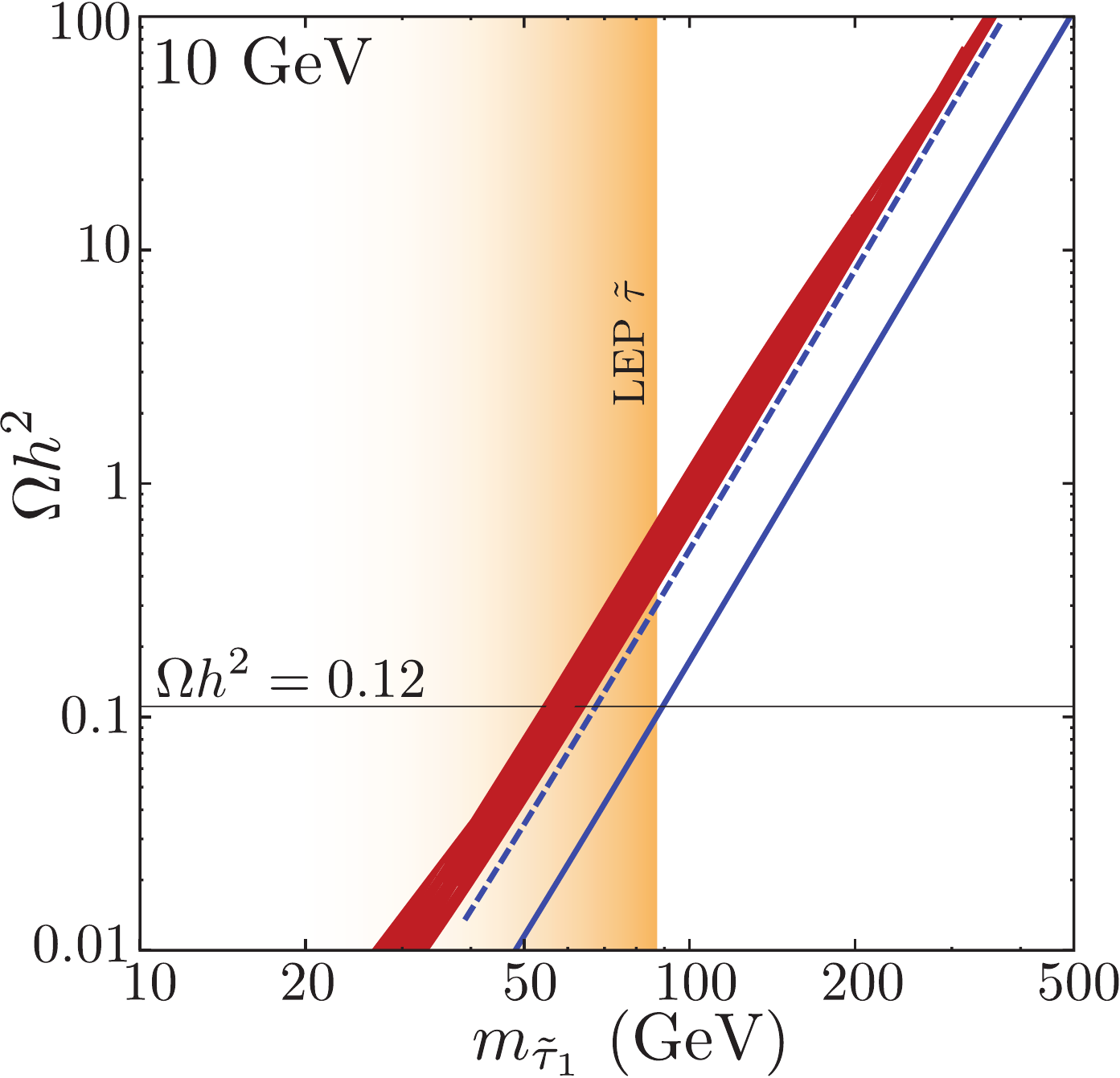} 
\includegraphics[width=2.9in]{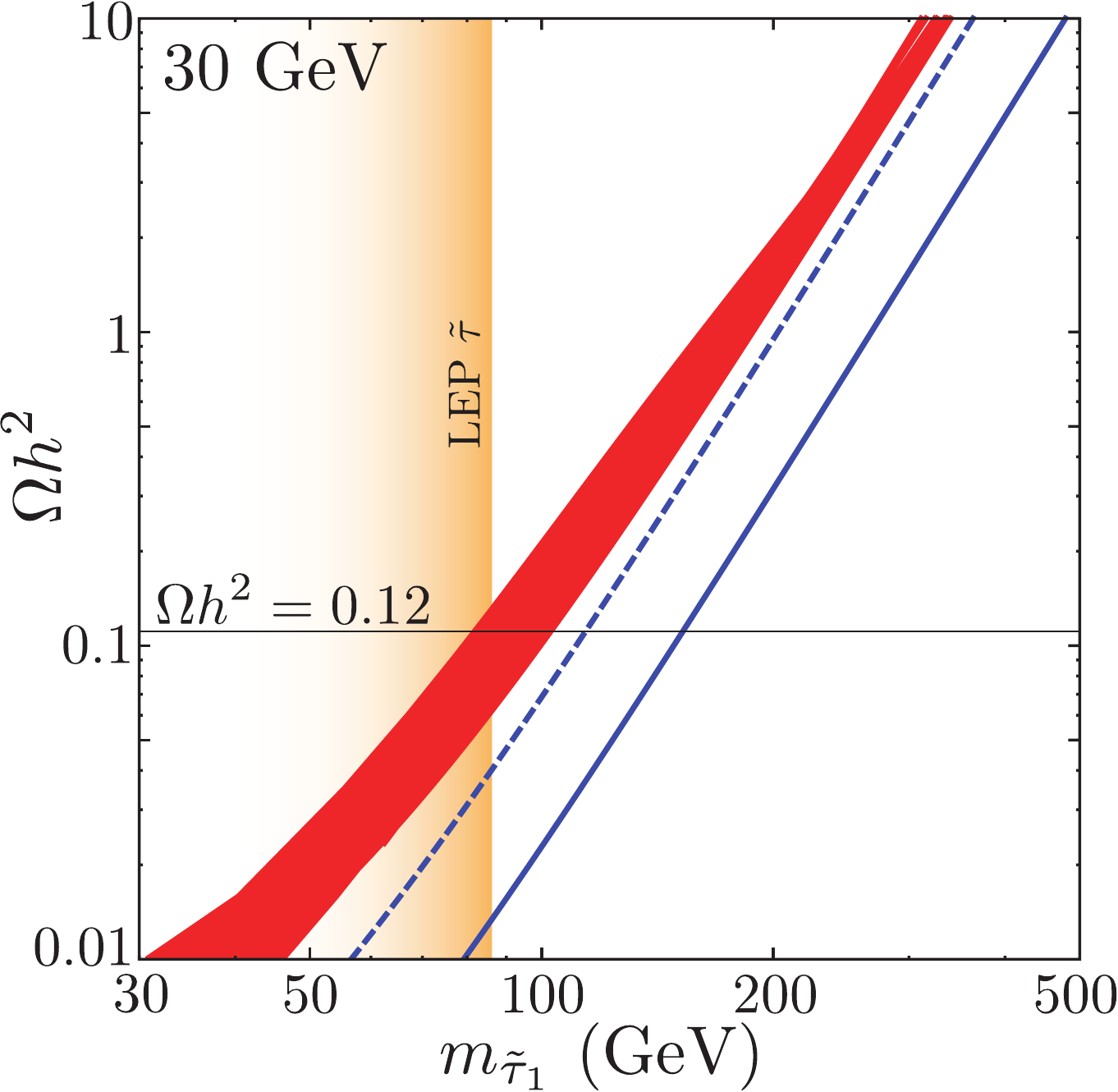} \\
\vspace{0.4cm}
\includegraphics[width=2.9in]{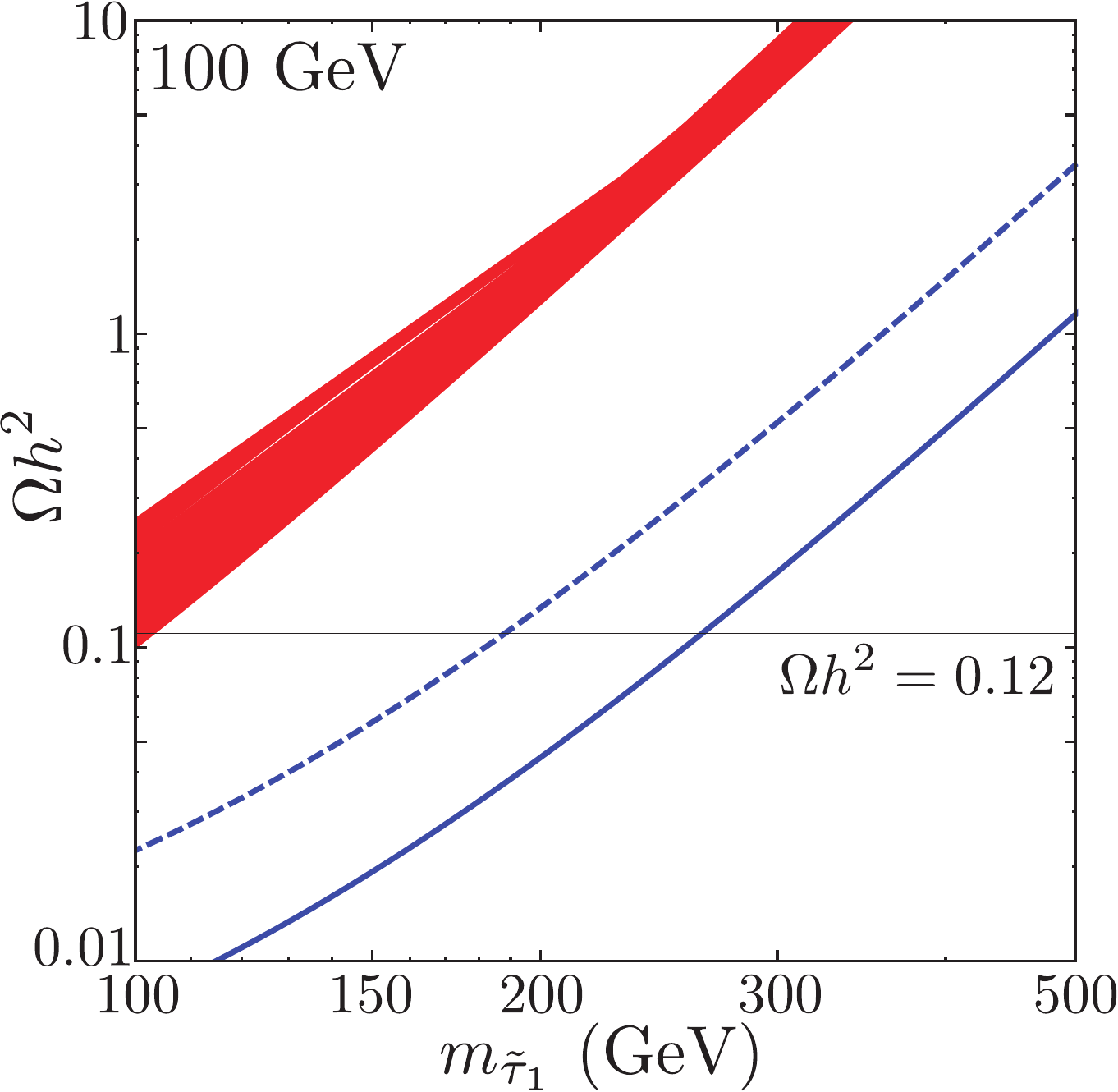} 
\includegraphics[width=2.9in]{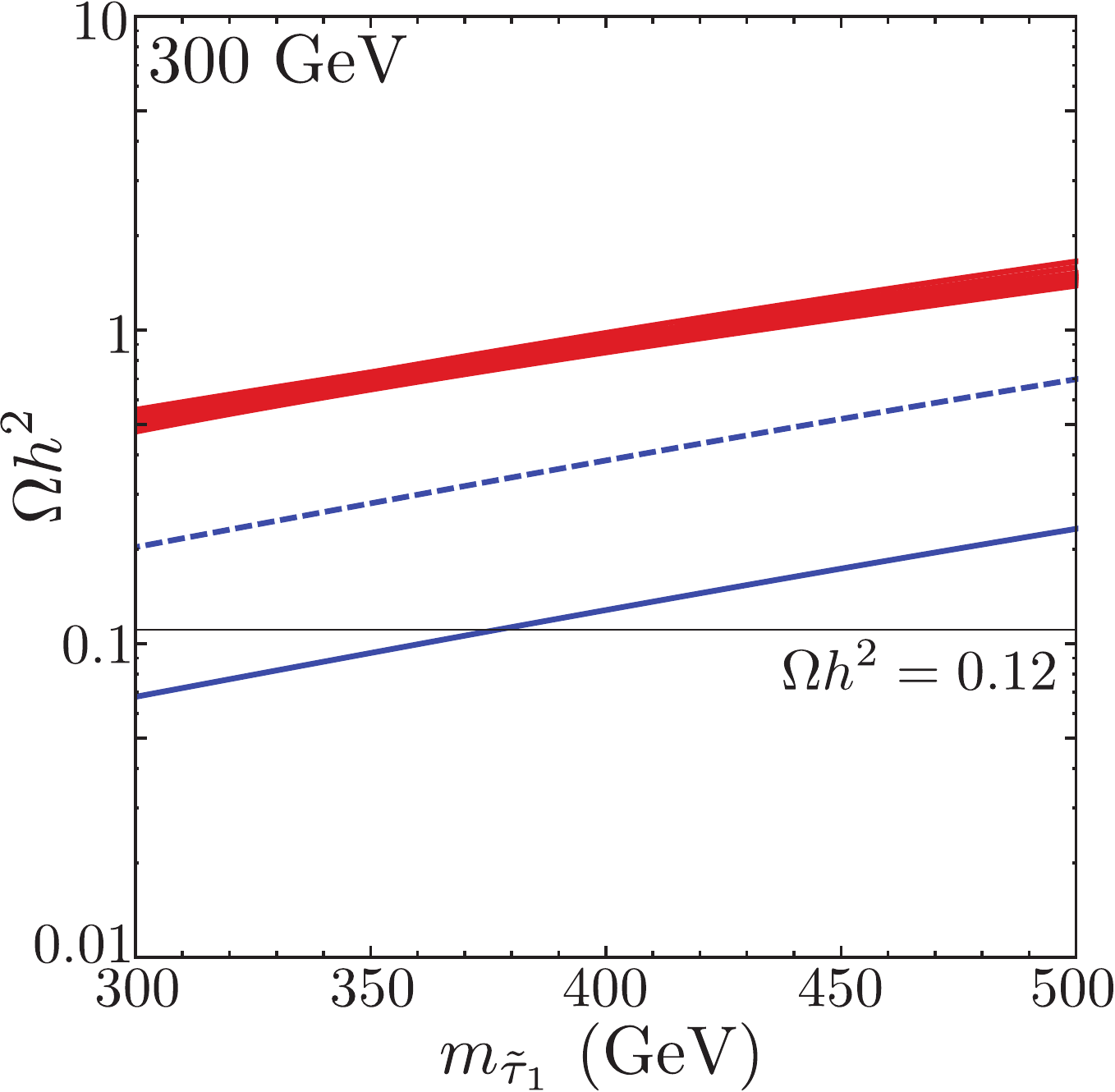} \\
\vspace{0.4cm}
\includegraphics[width=2.9in]{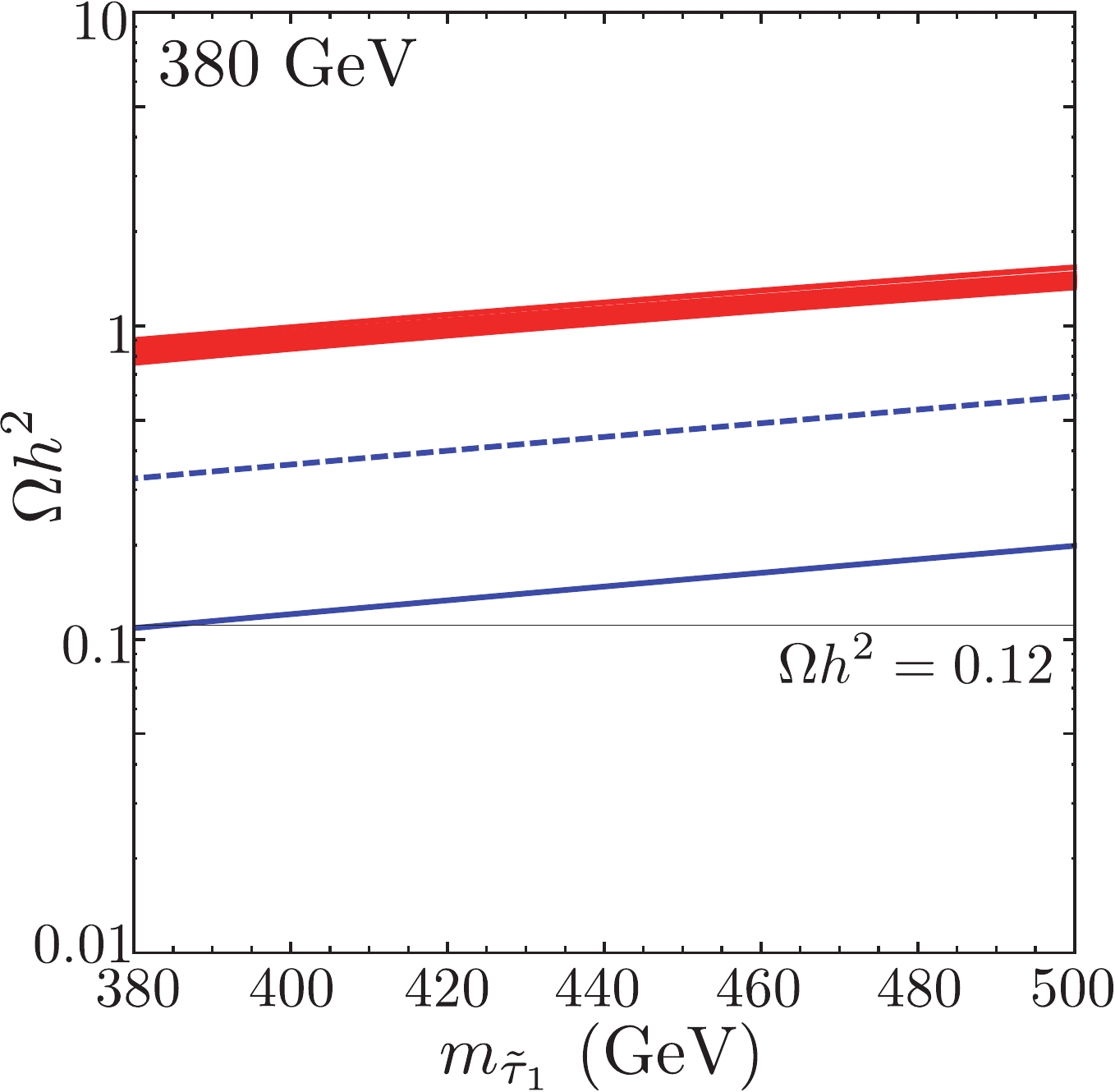} 
\caption{The thermal relic abundance of a Dirac bino, for several values of $m_{\chi}$ (10, 30, 100, 300 and 380 GeV), as a function of the lightest stau mass. The solid lines acount for annihilations through all three families of charged sleptons and sneutrinos, while the dashed lines include only the staus and tau sneutrinos. The red bands represent the predictions for a Majorana bino, over a wide range of MSSM parameters. For a Dirac bino with a mass in the range of $m_{\chi} \approx 10-380$ GeV, the observed dark matter abundance can be accommodated. No coannihilations have been included in these calculations. See text for details.} 
\label{diracrelic}
\end{figure*}

The annihilation cross section for any one of these processes is given by:
%
%\begin{widetext}
\begin{eqnarray}
&&\frac{d\sigma_{\chi \chi \rightarrow f\bar{f}}}{d \cos \theta} v_{\rm rel} = \frac{N_c \, \vec{k}}{64 \pi E^3} \bigg[\frac{1}{4}\sum_{\rm spins}|{\cal M}|^2\bigg] \\
& = & \frac{N_c}{64 \pi (m^2_{\chi}+\vec{p}^2)} \bigg[\frac{1}{4}\sum_{\rm spins}|{\cal M}|^2\bigg] \sqrt{1-\frac{m^2_f}{m^2_{\chi}+\vec{p}^2}}, \nonumber
\end{eqnarray}
%\end{widetext}
%
where $N_c=1$ (3) for annihilations to leptons (squarks). 

In Fig.~\ref{diracrelic}, we plot the thermal relic abundance of a bino-like Dirac neutralino annihilating through slepton exchange, for several values of the mass of the lightest neutralino (10, 30, 100, 300, and 380 GeV), and as a function of the lightest stau mass. In each frame, the solid line denotes the result including all three flavors of charged sleptons and sneutrinos, with $M_2=1.1 M_1$.  The dashed lines, in contrast, include only annihilations mediated by staus and tau sneutrinos. Also shown in each frame of Fig.~\ref{diracrelic} is the range of relic abundances predicted for a bino-like Majorana neutralino in the MSSM. To generate these bands, we varied the quantity $(A_{\tau}+\mu \tan \beta)$ betweeen 450 and $10^5$ GeV, and the ratio $M_L/M_R$ between 1.1 and 5. Recall that, as there is no left-right sfermion mixing in $R$-symmetric models, $\alpha=0$ in the Dirac case.

Searches at LEP and the LHC have provided lower limits for the various slepton masses relevant for these annihilation processes. LEP restricts staus to masses greater than 86 GeV, and smuons and selectrons to 100 GeV or higher (assuming, in each case, that sleptons decay to a lepton of the same flavor and a light neutralino)~\cite{pdg}. The stau constraint is shown in Fig.~\ref{diracrelic}. The LHC is most sensitive to left-handed smuons and selectrons, which they currently constrain to masses heavier than approximately 300 GeV~\cite{lhcslepton}. For right-handed sleptons, which provide the largest contributions to the neutralino's annihilation cross section (due to their larger hypercharge), the current bounds are much weaker~\cite{lhcsleptonrh}.

From Fig.~\ref{diracrelic}, we see that Dirac bino-like neutralinos always annihilate more efficiently than their Majorana counterparts. And whereas Majorana binos can only yield the desired thermal relic abundance for masses in the range of $m_{\chi} \sim 25-100$ GeV, we can find viable parameter space in the Dirac case for a much wider range of masses, $m_{\chi} \sim 10-380$ GeV. Note that in these calculations we have not included coannihilations between the LSP and sleptons, which could lead to significantly lower relic abundances if their masses are degenerate to within approximately $\sim$10\% or less.

\begin{figure}
\includegraphics[width=3.2in]{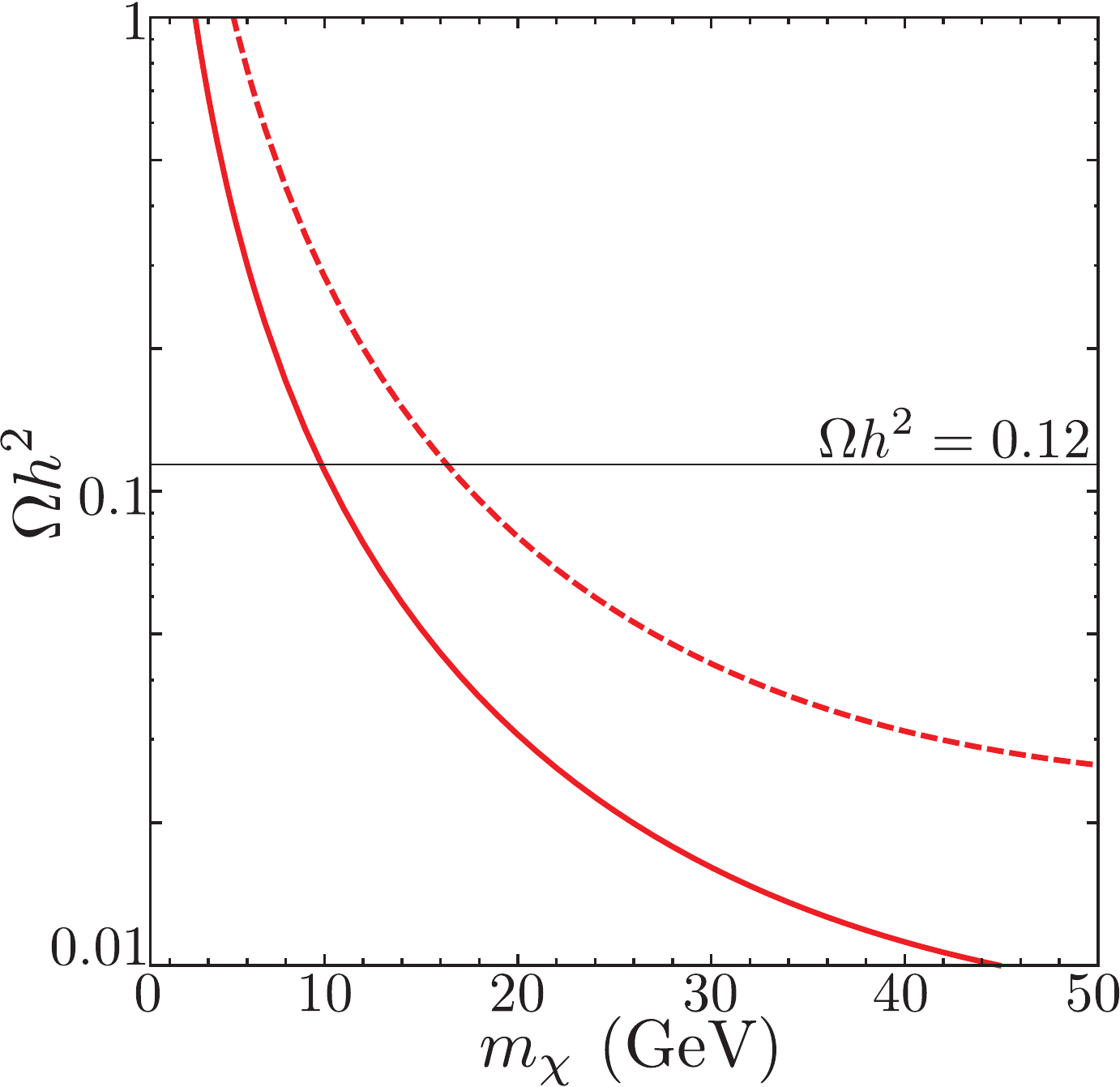} 
\caption{The thermal relic abundance of a Dirac bino as a function of $m_{\chi}$, with all slepton masses set to their minimum values allowed by accelerator constraints. Again, the solid and dashed lines include all three slepton generations and the tau generation only, respectively. See text for details.} 
\label{minslepton}
\end{figure}

In Fig.~\ref{minslepton} we show the relic abundance obtained for a Dirac neutralino, with all slepton masses set to their minimum allowed values. In particular, we have set both stau masses to 86 GeV, the right-handed smuon and selectron masses to 100 GeV, and the left-handed smuon and selectron masses to 300 GeV. Again, the solid and dashed lines include all three generations of sleptons, and only third generation sleptons, respectively. One should also keep in mind that if any of the bino-lepton-slepton couplings are flavor violating, lighter sleptons masses could be possible, allowing for even higher annihilation cross sections and lower predictions for the thermal relic abundance.

For a Majorana bino-like neutralino, the annihilation cross section to leptons through slepton exchange can be constrained by measurements of the muon and electron magnetic moments. In $R$-symmetric models, however, such constraints are negligible.  This is because the contributions to $(g-2)$ do not benefit from left-right slepton mixing in $R$-symmetric models, and thus require a lepton helicity-flip, suppressing the amplitude by a factor of $\sim m_l/m_{\chi}$.

\section{Implications for Indirect Detection}
\label{indirect}

In the previous section, we found that a Dirac bino-like neutralino will annihilate largely to charged lepton pairs. If the stau, smuon and selectron masses are approximately equal, we expect Dirac neutralinos to annihilate to roughly equal numbers of $\tau^+ \tau^-$, $\mu^+ \mu^-$ and $e^+ e^-$ final states, with a low-velocity annihilation cross section (per generation) given by:
\begin{eqnarray}
\sigma_{l^+ l^-} v &=& \frac{g'^4 \, m^2_{\chi}}{8 \pi} \bigg[ \frac{1}{(m^2_{{\tilde l}_R}+m^2_{\chi})^2} + \frac{1}{16\,(m^2_{{\tilde l}_L}+m^2_{\chi})^2} \bigg] \nonumber \\
&\approx& 1.3\times 10^{-26} \, {\rm cm}^3/{\rm s} \, \bigg(\frac{m_{\chi}}{100 \, {\rm GeV}}\bigg)^2 \nonumber \\
&\times& \bigg(\frac{(100\,{\rm GeV})^2+(250 \, {\rm GeV})^2}{m^2_{{\tilde l}}+m^2_{\chi}}\bigg)^2.
\end{eqnarray}
We also note that as the terms of the annihilation cross section proportional to velocity squared ($\sigma v \propto v^2$) are opposite in sign to the low-velocity terms, a mild cancellation occurs in the relic abundance calculation. As a result, models which predict the desired thermal relic abundance have a low-velocity cross section that is approximately 20\% larger than the canonical expectation for a simple relic (20\% larger than $\simeq 3\times 10^{-26}$ cm$^3$/s), somewhat enhancing the prospects for indirect detection.

These results are in stark contrast to the dominant (low-velocity) annihilation channels predicted for Majorana neutralinos, which consist almost entirely of heavy fermions ($b\bar{b}$, $t \bar{t}$, $\tau^+ \tau^-$), and/or combinations of gauge and Higgs bosons ($W^+ W^-$, $ZZ$, $ZH$, $Zh$, $W^{\pm} H^{\mp}$, $HA$, $hA$).  The prediction of annihilations to leptonic and roughly flavor-democratic final states has a number of significant implications for indirect searches:
\begin{itemize}
\item{Annihilations to $e^+ e^-$ are predicted to lead to a distinctive edge-like feature in the spectra of cosmic ray electrons and positrons, even after energy losses and other propagation effects are taken into account~\cite{Baltz:2004ie}. The lack of such a feature in the positron fraction measured by the AMS experiment significantly constrains such annihilations. For example, an annihilation cross section of $10^{-26}$ cm$^3$/s to $e^+ e^-$ requires $m_{\chi} \gsim 60$ GeV~\cite{Bergstrom:2013jra}.}
\item{Constraints from measurements of cosmic ray antiprotons are very weak for Dirac neutralinos, as annihilations to leptonic channels do not contribute to this signal.}
\item{The gamma-ray spectrum produced from Dirac neutralino annihilations is dominated by the decays of tau leptons, which yield a much harder spectrum than most other annihilation channels. Constraints from observations of dwarf spheroidal galaxies~\cite{GeringerSameth:2011iw,Ackermann:2011wa} and the Galactic Center~\cite{Hooper:2012sr} from the Fermi gamma-ray space telescope limit $m_{\chi} \gsim 10$ GeV  for $\sigma v \simeq 10^{-26}$ cm$^3$/s to $\tau^+ \tau^-$.}
\item{Annihilations to $\tau^+ \tau^-$ and $\nu \bar{\nu}$ final states taking place in the core of the Sun could produce a significant flux of high-energy neutrinos, especially in light of the large elastic scattering cross sections with nuclei potentially expected for a Dirac neutralino.  Althought current limits do not seem to constrain Dirac neutralinos~\cite{Aartsen:2012kia,Kappl:2011kz}, future large-volume, low-threshold experiments may be sensitive to such particles, especially in the low-mass range, which we will discuss further in the following section.}
\end{itemize}

\section{Light Dirac Neutralinos and Recent Direct and Indirect Detection Anomalies}
\label{anomalies}

In recent years, a number of direct detection experiments have reported results which can be interpreted as possible indications of dark matter scattering~\cite{Aalseth:2012if,Aalseth:2011wp,Agnese:2013rvf,Collar:2012ed,Angloher:2011uu,damanew,xenondan}.
 Collectively, these signals favor a dark matter particle with a mass of $\sim$7-10 GeV and a spin-independent scattering cross section of $\sim 2\times 10^{-41}$ cm$^2$. In addition, a spatially extended excess of gamma-rays observed from the region of the Galactic Center and throughout the Inner Galaxy can be explained by a dark matter particle with a similar mass, annihilating to $\tau^+ \tau^-$, possibly among other leptons~\cite{gc1,gc2,gc4,gc5,newgcchris}. While the large elastic scattering cross section and leptonic annihilation channels implied by these observations are not typically exhibited by neutralinos within the context of the MSSM, Dirac neutralinos in $R$-symmetric models can much more easily account for these signals.

To generate the spin-independent elastic scattering cross section required to explain the reported direct detection anomalies, we must consider a light neutralino with a small, but not insignificant, higgsino fraction, $|N_{13}|^2 - |N_{14}|^2 \sim 0.02$ (corresponding to $\mu_u$ or $\mu_d \sim 300$ GeV).  Although the coupling of a light neutralino to the $Z$ is constrained by LEP's measurement of the invisible decay width, $\Gamma_{Z\rightarrow {\rm inv}}$~\cite{pdg}, the coupling required to accommodate these anomalies is consistent with this result ($\Gamma_{Z \rightarrow \chi \chi} < 3$ MeV at the 95\% confidence level, which in the case of a Dirac neutralino corresponds to $|N_{13}|^2 -|N_{14}|^2 \lsim 0.06$)\cite{Fitzpatrick:2010em}. 

% for 2x10^-41, need sigma_n ~5 x10^-41, 

% \mu = 10*0.938 / 10.938 --> factor of 0.836
% need (N13^2 -N14^2 )^2 /0.01^2 = 2.3 x 10
%  |N_{13}|^2 - |N_{14}|^2 \approx 0.048 

In order for a $\sim$10 GeV Dirac neutralino to avoid exceeding the constraints from AMS~\cite{Bergstrom:2013jra}, we must suppress the annihilation cross section to $e^+ e^-$ by increasing the mass of the selectrons to $m_{\tilde e} \gsim 200$ GeV. In this case, annihilations will proceed largely to taus and/or muons. If the stau and smuon masses are not far above the minimum values allowed by LEP and the LHC, these annihilation channels could potentially accommodate the observed gamma-ray excess from the Galactic Center and Inner Galaxy, while also generating an acceptable thermal relic abundance.

\section{Dirac or Pseudo-Dirac?}
\label{pseudodirac}

\begin{figure*}
\includegraphics[width=3.5in]{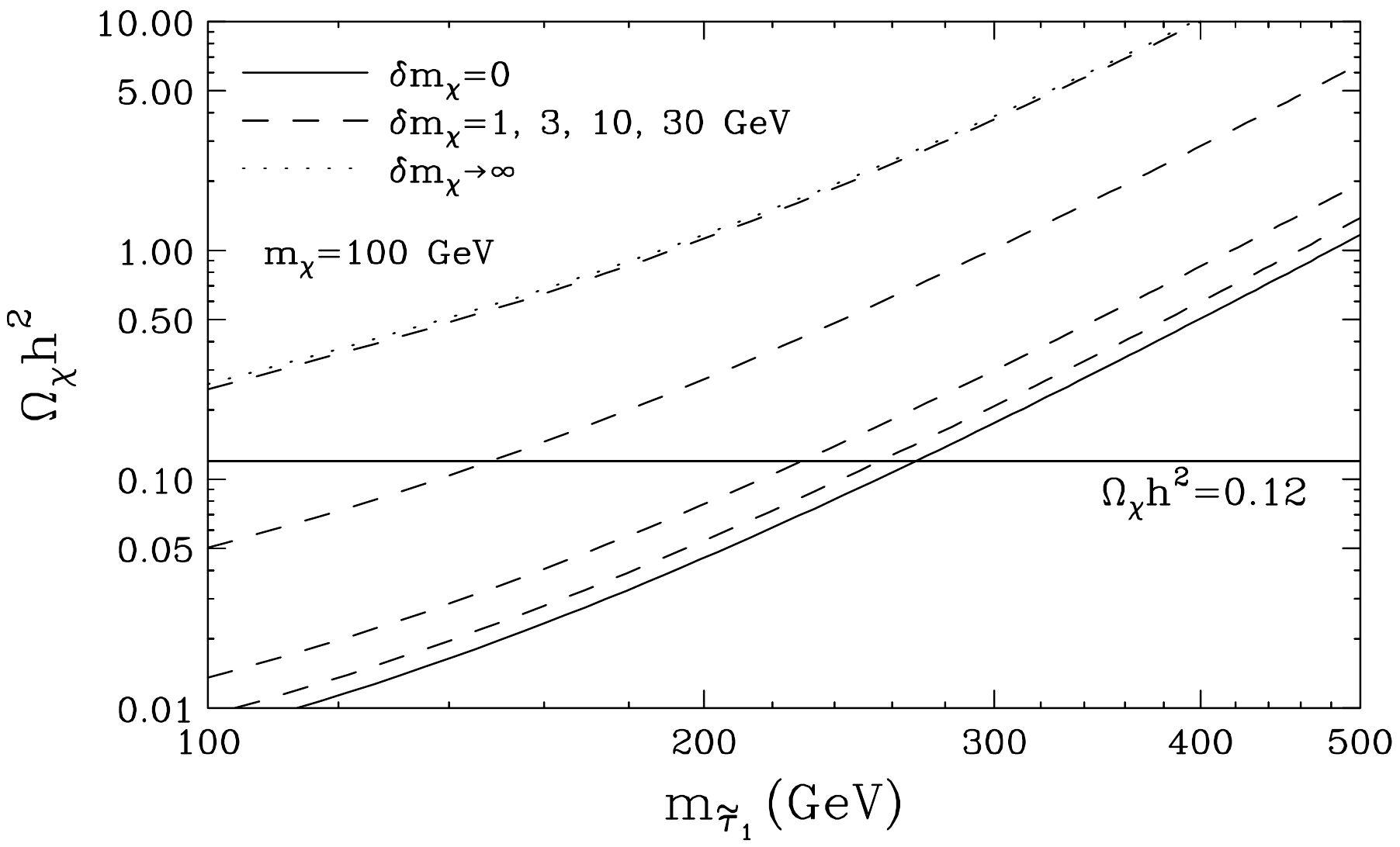} 
\includegraphics[width=3.5in]{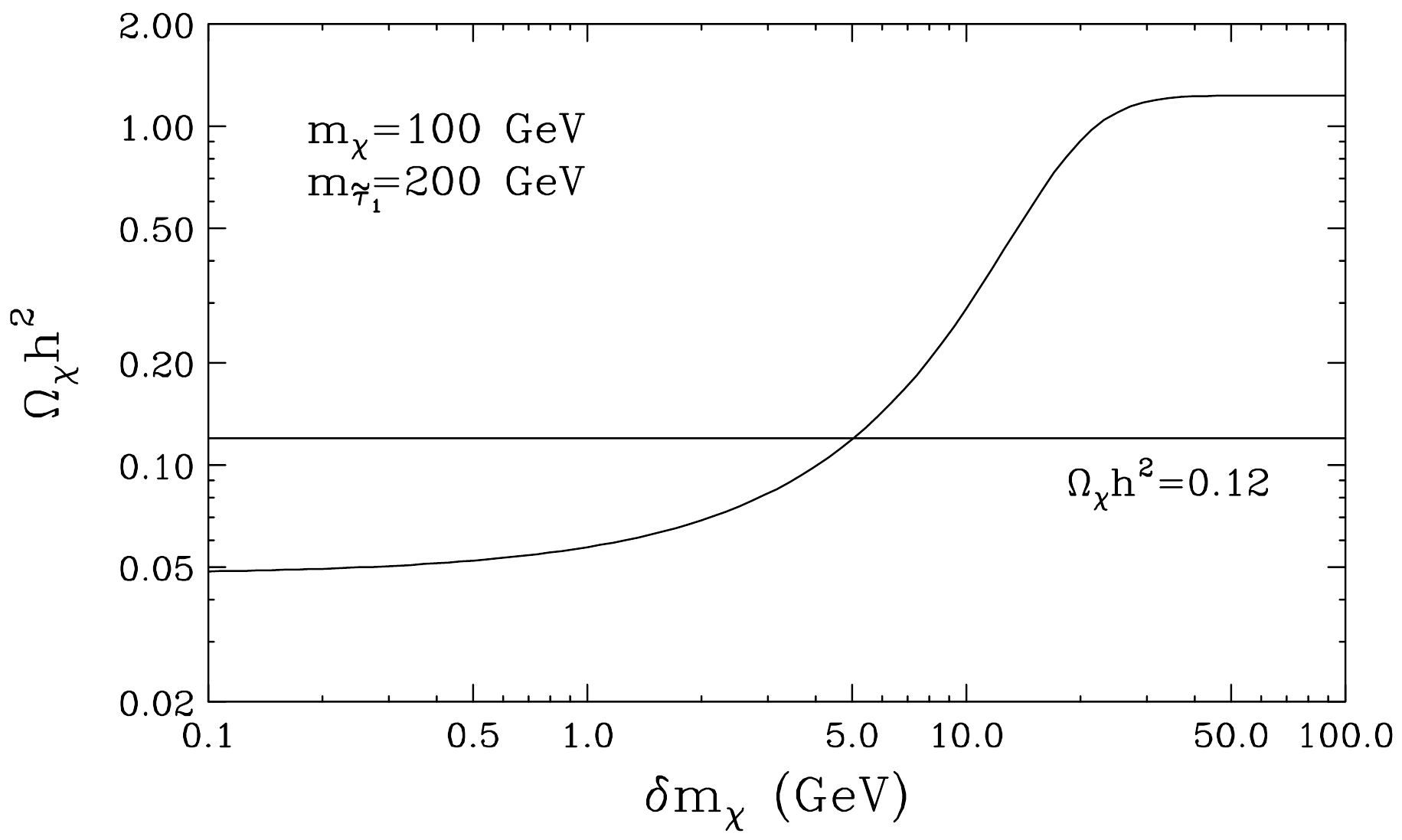} 
\caption{Left: The thermal relic abundance of a 100 GeV pseudo-Dirac bino, for several values of the mass splitting between the two Majorana fermions that make up the pseudo-Dirac state, $\delta m_{\chi}=$ 0, 1, 3, 10, 30 GeV, as a function of the lightest stau mass. Right: The thermal relic abundance as a function of the mass splitting. For $\delta m_{\chi} \ll T_{\rm FO}$, the relic abundance is the same as in the pure Dirac case, while for $\delta m_{\chi} \gsim T_{\rm FO}$, the Majorana-like behavior is recovered. See text for details.} 
\label{pdrelic}
\end{figure*}

Up to this point in our study, we have explicitly assumed that $R$-symmetry is unbroken, and thus that our neutralino dark matter candidate is a Dirac particle, with no Majorana mass terms.  This need not be the case, however, as small Majorana mass terms could slightly break the $R$-symmetry~\cite{Zur:2008zg,Benakli:2008pg} without spoiling many of the attractive features motivating such models~\cite{Kribs:2012gx,Kribs:2007ac,Fok:2010vk,Fox:2002bu}.  And although global supersymmetry breaking without breaking $R$-symmetry is not difficult to imagine (and may, in fact, be preferable~\cite{Murthy:2007qm}), adjustments to the superpotential of supergravity which allow for the suppression the cosmological constant appear to introduce a degree of $R$-symmetry violation~\cite{Dine:1992yw}. And while such $R$-symmetry violation may originate within a hidden sector, anomaly mediation is expected to transfer it to the low-energy sparticle spectrum, spliting the Dirac neutralino into two Majorana states with a mass difference of the following order~\cite{Bagger:1999rd}:
\begin{eqnarray}
\delta m_{\chi} &\sim& \frac{\alpha \, m_{3/2}}{4 \pi} \\ \nonumber
&\sim& 60 \,\, {\rm MeV} \times \bigg(\frac{m_{3/2}}{100 \, {\rm GeV}}\bigg) 
\end{eqnarray}
where $m_{3/2}$ is the mass of the gravitino. As we will show, such a mass splitting would restore the Majorana-like nature of the LSP for the purposes of both direct and indirect detection (although not necessarily for the purposes of the relic abundance calculation).  Bearing in mind proposals for how such mass terms might be highly suppressed~\cite{Bagger:1999rd,Luty:2002ff,Luty:2002hj,ArkaniHamed:2004yi}, and given our limited understanding of the cosmological constant problem, we remain agostic about whether $R$-symmetry is likely to be a broken or unbroken symmetry of nature.  In this section we consider calculations of the relic abudance, elastic scattering, and indirect detection in the case in which the lightest neutralino is a pseudo-Dirac state of two quasi-degenerate Majorana fermions.

\subsection{Relic abundance of a pseudo-Dirac neutralino}

We begin with an expression for the effective annihilation cross section, accounting for annihilations and coannihilations between the two quasi-degenerate Majorana states, written as a function of $x \equiv T/m_{\chi}$~\cite{Servant:2002aq}:
\begin{widetext}
\begin{equation}
%\sigma_{\rm Eff} v (x) = \sum_{i,j} \sigma_{ij} v \, \frac{g_i g_j}{g^2_{\rm Eff}} \, (1+\Delta_i)^{3/2} (1+\Delta_j)^{3/2} \exp[-x\,(\Delta_i+%\Delta_j)],
%
\sigma_{\rm Eff} (x) = \frac{4}{g^2_{\rm Eff}}\bigg[\sigma_{11}  + 2 \, \sigma_{12} \, (1+\Delta)^{3/2} \exp[-x\, \Delta]+\sigma_{22} \, (1+\Delta)^{3} \exp[-2\,x\,\Delta] \bigg],
\end{equation}
\end{widetext}
 where $\Delta=(m_{\chi^2}-m_{\chi^1})/m_{\chi^1}$ is the fractional mass splitting between the two Majorana states. The effective number of degrees of freedom is defined by:
\begin{equation}
g_{\rm Eff} (x) = 2+ 2 \, (1+\Delta)^{3/2} \, \exp[-x \, \Delta].
\end{equation}
The thermal average of the cross section is found by integrating over the thermal history surrounding freeze-out:
\begin{equation}
I_a = x_F \int^{\infty}_{x_F} a_{\rm Eff} \, x^{-2} dx, \,\,\,\,\,\,\,\,\,\,\,\,\,\, I_b = 2 x^2_F \int^{\infty}_{x_F} b_{\rm Eff} \, x^{-3} dx,
\end{equation}
where $\sigma_{\rm Eff} v = a_{\rm Eff} + b_{\rm Eff} v^2$. In terms of these quantities, the thermal abundance of dark matter is given by
\begin{eqnarray}
\Omega_{\chi} h^2 &\approx& \frac{1.04\times10^9}{M_{\rm Pl}} \frac{x_F}{\sqrt{g_{\star}}} \frac{1}{(I_a+3I_b/x_{F})} \nonumber \\
&\approx& 0.12 \, \bigg(\frac{3\times 10^{-26} \, {\rm cm}^3/{\rm s}}{I_a+3I_b/s_F}\bigg).
\end{eqnarray}
If $x \, \Delta \gg 1 $ (or equivalently, if $\delta m_{\chi} \gg T_{\rm FO}$), then the $\sigma_{12}$ and $\sigma_{22}$ contributions become exponentially suppressed, and $\sigma_{\rm Eff} \approx \sigma_{11}$. If the mass splitting is comparable to or smaller than the freeze-out temperature, however, these additional terms can play an important role. The $\sigma_{12}$ term, in particular, allows for efficient $s$-wave annihilation to light fermion final states, without the $(m_f/m_{\chi})^2$ suppression that is exhibited by Majorana dark matter candidates. 

In Fig.~\ref{pdrelic}, we show how the thermal relic abundance of a pseudo-Dirac bino depends on the mass splitting between the two Majorana states. We find that for mass splittings much smaller than the freeze-out temperature, the resulting relic abundance is the same as in the pure Dirac case. In contrast, if $\delta m_{\chi} \gsim T_{\rm FO}$, the lightest neutralino freezes out like an isolated Majorana state.

\subsection{Direct and indirect detection of a pseudo-Dirac neutralino}

\begin{figure*}
\includegraphics[width=4.5in]{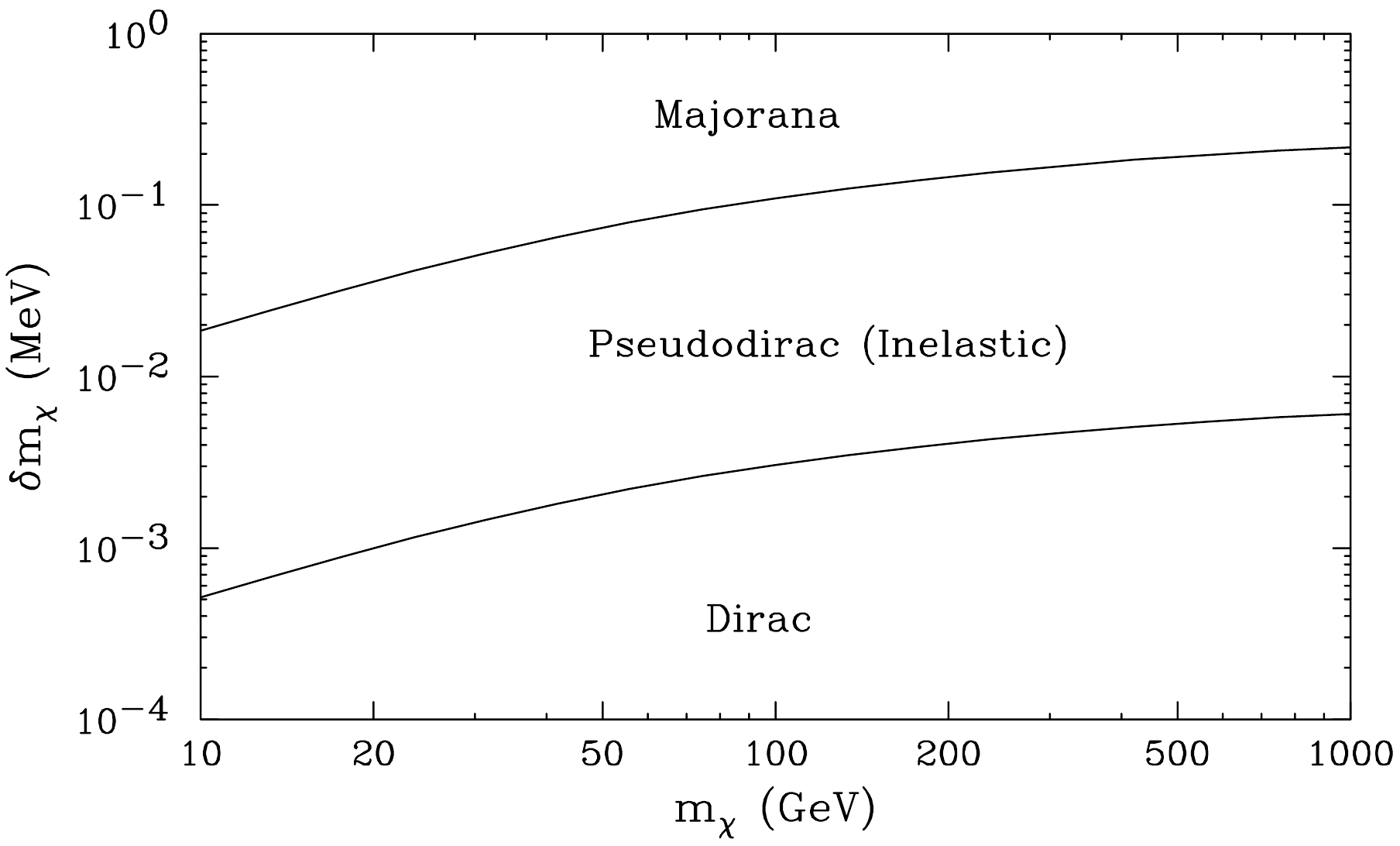}
\caption{Depending on the mass splitting between the two states which together constitute our Dirac neutralino dark matter candidate, it may scatter with nuclei effectively as a single Dirac state, or scatter only inelastically as a pseudo-Dirac state, or as an isolated Majorana state. For a xenon target, we show the thresholds for these scenarios, corresponding to dark matter velocities of 100 km/s (lower) and 600 km/s (upper).} 
\label{inelastic}
\end{figure*}

If there is even a very small splitting between the masses of the two Majorana fermions, the dark matter present in the Universe today will be overwhelmingly dominated by the lightest of these states, as the lifetime of the heavier state will almost certainly be short compared to the age of the Universe.  To estimate the lifetime for $\chi_2 \rightarrow \chi_1$, we consider the magnetic moment operator:
\begin{equation}
{\cal L} \supseteq \frac{e^3}{16\pi^2} \frac{\delta m_{\chi}}{\tilde{m}^2} \bar{\chi}_2 \sigma^{\mu\nu} \chi_1 F_{\mu\nu},
\end{equation}
where $\tilde{m}$ is the mass of the slepton running in the loop, and the prefactor is estimated based on a loop-suppression, three powers of the electromagnetic coupling, and then dimension-counting. The width of the decay is then given by:
\begin{eqnarray}
\Gamma &\approx& \frac{1}{16\pi} \frac{\delta m_{\chi}}{m_{\chi}^2} |{\cal M}|^2 \\
 %               &= &\frac{1}{16 \pi} \frac{\delta m_{\chi}}{m_{\chi}^2} \frac{e^6}{(16 \pi^2)^2} \frac{\delta m_{\chi}^2}{\tilde{m}^4} (128 \, m_{\chi}^3) \\ 
                &\approx & 2 \alpha^3 \frac{\delta m^3_{\chi}}{\tilde{m}^4} m_{\chi}^2 \nonumber
\end{eqnarray}
which leads to a lifetime of
\begin{equation}                
\Gamma^{-1} = 1.4 \times 10^{-4}~\mbox{s} \left(\frac{1~\mbox{MeV}}{\delta m_{\chi}}\right)^3 \left(\frac{\tilde{m}}{200~\mbox{GeV}}\right)^4 \left( \frac{100~\mbox{GeV}}{m_{\chi}}\right)^2.
\end{equation}
Therefore, only for sub-eV mass splittings will the lifetime for this process be comparable to or longer than the age of the Universe. For direct and indirect detection, this has important implications:
\begin{itemize}
\item{If the kinetic energy in a neutralino-nuclei scattering event is unable to exceed the mass difference between the two Majorana states, the (spin-dependent) scattering will proceed as if the neutralino were an isolated Majorana state. If the mass splitting is more modest, however, the lightest neutralino may scatter inelastically and spin-independently, transforming in the interaction into the slightly heavier state~\cite{inelastic}. In Fig.~\ref{inelastic}, we show where the approximate boundaries for these regimes lie, for the case of scattering with xenon nuclei. For the upper and lower boundaries shown, we have taken velocities which approximately bracket those present in the Galactic Halo, 600 km/s and 100 km/s, respectively.}
\item{Without the presence of a significant population of the heavier state, neutralinos annihilating in the Galactic halo will behave as Majorana particles, with the standard MSSM-like suppression for annihilations to light fermions.}
\end{itemize}

To summarize this section, a pseudo-Dirac neutralino will maintain some elements of its Dirac-like nature, depending on the size of the mass splitting.  For $\delta m_{\chi} \gsim$~eV, the neutralinos behave like Majorana particles for the purposes of indirect detection, while direct detection retains much of its Dirac-like features for splittings as large as $\delta m_{\chi} \sim$1-100 keV. For the calculation of the thermal relic abundance, small splittings have little impact. Only for mass splittings larger than the freeze-out temperature does the behavior significantly depart from that predicted in the pure-Dirac case.

%%%%

\section{Summary and Conclusions}
\label{summary}

In this paper, we have discussed the dark matter phenomenology of Dirac neutralinos, as predicted in supersymmetric models with an unbroken $R$-symmetry. Such scenarios are theoretically and phenomenologically well motivated, and lead to a dark matter candidate with many features that are very different from those exhibited by the Majorana neutralinos found within the MSSM and most other supersymmetric models. In particular:
\begin{itemize}
\item{Dirac neutralinos undergo coherent (spin-independent) scattering with nuclei as a result of vector interactions with the $Z$ and with squarks. To evade current constraints from XENON100 and other direct detection experiments, the lightest neutralino must either be highly bino-like (with little higgsino mixing) and the squarks must be quite heavy ($m_{\tilde q} \gsim 2$ TeV), or the lightest neutralino must be rather light, $m_{\chi} \lsim 20$ GeV.}
\item{A bino-like Dirac neutralino with a mass in the range of 10-380 GeV can annihilate through slepton exchange to generate a thermal relic density consistent with the observed dark matter abundance, without relying on coannihilations or annihilations near a resonance. Dirac neutralinos do not experience the chirality suppression predicted for Majorana dark matter candidates when annihilating into light fermions.}
\item{Dirac bino-like neutralinos annihilate largely to charged leptons pairs, leading to indirect detection signatures which are very different from those predicted for Majorana dark matter candidates. At present, the strongest indirect detection constraints come from the lack of spectral features observed in AMS's measurement of the cosmic ray positron fraction.}
\item{A number of recent direct and indirect detection anomalies, potentially interpretable as signals of $\sim$7-10 GeV dark matter particles, can be accommodated by a Dirac neutralino. In particular, the large elastic scattering cross section and leptonic annihilation channels predicted in this model provide a good match to these signals.}
\item{If the $R$-symmetry is slightly broken, the Dirac neutralino may be split into two quasi-degenerate Majorana states. In this pseudo-Dirac case, some or all of the Dirac-like nature of the dark matter phenomenology can be altered, depending on the size of the mass splitting.}
\end{itemize}

\bigskip
%\newpage

{\it Acknowledgements}:  The authors are grateful to the Aspen Center for Physics, the Kavli Center for Theoretical Physics, the University of Utah, the Center for Theoretical Underground Physics and Related Areas (CETUP 2013), and the organizers of TeVPA 2012,  for their support and hospitality during the completion of this work. We would like to thank Claudia Frugiuele and Paddy Fox for valuable discussions. This work has been supported in part by the US Department of Energy, including grants DE-FG02-04ER-41291 and DE-FG02-13ER-41913.

\end{document}